\documentclass[superscriptaddress, amsmath,amssymb, aps, twocolumn,nofootinbib]{revtex4-1}

\usepackage{graphicx}
\usepackage{dcolumn}
\usepackage{bm}
\usepackage{wasysym}
\usepackage{hyperref}
\usepackage[mathlines]{lineno}
\usepackage{xcolor}
\usepackage{dsfont}

\usepackage{gensymb}
\usepackage{enumitem}
\usepackage{multirow}
\usepackage{booktabs}
\usepackage{array}
\newcolumntype{P}[1]{>{\centering\arraybackslash}p{#1}}
\usepackage{amssymb}
\usepackage{adjustbox}
\usepackage{orcidlink}
\usepackage{float}
\usepackage{tikz}
\usetikzlibrary{positioning}
\usetikzlibrary{calc}

\usepackage[normalem]{ulem}

\newcommand*{\defeq}{\mathrel{\vcenter{\baselineskip0.5ex \lineskiplimit0pt
                     \hbox{\scriptsize.}\hbox{\scriptsize.}}}
                     =}

\newcommand{\iu}{\mathrm{i}\mkern1mu}
\newcommand{\du}{\mathrm{d}}

\newcounter{count}

\begin{document}

\title{Black Hole-Boson Star Binaries: Gravitational Wave Signals and Tidal Disruption}

\author{Gareth Arturo Marks \orcidlink{0009-0003-3160-9337}}
\email{gam54@cam.ac.uk}
\affiliation{DAMTP, Centre for Mathematical Sciences,
University of Cambridge, Wilberforce Road, Cambridge CB3 0WA, UK}

\author{Seppe J. Staelens \orcidlink{0000-0002-1262-1600}}
\email{ss3033@cam.ac.uk}
\affiliation{DAMTP, Centre for Mathematical Sciences,
University of Cambridge, Wilberforce Road, Cambridge CB3 0WA, UK}
\affiliation{Leuven Gravity Institute, KU Leuven,
Celestijnenlaan 200D box 2415, 3001 Leuven, Belgium}

% \author{Tamara Evstafyeva \orcidlink{0000-0002-2818-701X}}
% \email{tevstafyeva@perimeterinstitute.ca}
% \affiliation{Perimeter Institute for Theoretical Physics,
% Waterloo, Ontario N2L 2Y5, Canada}

\author{Ulrich Sperhake.
\orcidlink{0000-0002-3134-7088}}
\email{U.Sperhake@damtp.cam.ac.uk}
\affiliation{DAMTP, Centre for Mathematical Sciences,
University of Cambridge, Wilberforce Road, Cambridge CB3 0WA, UK}
\affiliation{Department of Physics and Astronomy, Johns Hopkins University,
3400 North Charles Street, Baltimore, Maryland 21218, USA}
\affiliation{{TAPIR 350-17, Caltech, 1200 E. California Boulevard,
Pasadena, California 91125, USA}}

\date{\today}

%=============================================================================
\begin{abstract}
We present a detailed, fully nonlinear study of binary systems involving one black hole and one boson star, considering the effects of both a quartic self-interaction and a solitonic potential for the scalar field. First, we show the importance of using initial data for which the boson star is in an equilibrated configuration to obtain accurate gravitational waveforms, and discuss methods to further improve constraint violations in the initial data. We then present a series of head-on collisions, showing that even in this simplified scenario the radiative efficiency varies significantly with the scalar potential chosen. In addition to this, we present a preliminary study of inspiral configurations, showing that an appropriate scalar self-interaction can suppress tidal disruption. We comment throughout on implications for attempts to build model-agnostic waveform template banks for exotic compact objects.
\end{abstract}

\maketitle

%=============================================================================
\section{Introduction}
\label{sec:intro}

Since the first detection of gravitational waves (GWs) from a black hole (BH) binary merger \cite{Abbott:2016blz}, both the number and accuracy of GW events recorded have rapidly increased \cite{LIGOScientific:2018mvr,KAGRA:2021vkt, LVK_gwtc4}.
With next-generation detectors such as the Einstein Telescope \cite{abac2025sciencET}, Cosmic Explorer \cite{Evans:2023euw} and the Laser Interferometer Space Antenna (LISA) \cite{amaroseoane2017lisa} on the horizon, both the frequency bands and strain sensitivities available for exploration are set to further improve, making GW astronomy an increasingly promising avenue for the study of open problems in fundamental physics and cosmology.

A central question that GW astronomy aims to address is the \textit{black hole hypothesis}-- whether all the astrophysical objects believed to be BHs are indeed well-modeled by the classical BHs of general relativity. 
To date, all observations of GW events \cite{LIGOScientific:2021sio}, as well as interferometric observations of BH environments using the Event Horizon Telescope \cite{EventHorizonTelescope:2019dse, EventHorizonTelescope:2022wkp} are consistent with classical BHs.
However, there are still several ways this hypothesis could turn out to be incomplete and, hence, in need of testing: either modifications to classical general relativity could introduce novel effects into waveforms \cite{Evstafyeva:2022rve, Berti:2015itd, Okounkova:2017yby,Witek:2018dmd,Sperhake:2017itk}, 
or a population of exotic compact objects (ECO) could exist among, or even replace, classical BHs.
This latter possibility is particularly intriguing, given the ability of horizonless ultracompact objects (UCOs) ---
i.e. objects possessing a pair of light rings --- to mimic a broad range of BH phenomenology, including shadows \cite{Herdeiro_2021, sengo2024imitationgamereloadedeffective}, polarimetry imprints \cite{Rosa_2025} and ringdown spectra \cite{Synge:1966okc, Siemonsen:2024snb}, all of which are closely related to the light-ring structure of the object \cite{Cardoso:2008bp,Koga:2022dsu,Volkel:2022khh}.
While UCOs may suffer from a dynamical instability that, if efficient,  would limit their viability
as astrophysical BH mimickers \cite{Cardoso:2014sna}, recent nonlinear dynamical evolutions of UCOs have found evidence that the instability may not be generally effective \cite{Marks_2025, Marks_2025_CP, Staelens:2025wom, Evstafyeva:2025mvx}, adding further credence to their viability as
physical objects.
Exploring the prospects for an ECO population existing in the Universe requires a detailed understanding of how GW signals from mergers involving ECOs (and particularly UCOs) deviate  from those emitted by standard BH binaries.

A conceptually simple yet phenomenologically rich ECO model is given by boson stars (BSs), which are self-gravitating stationary configurations of a fundamental bosonic field  \cite{Kaup:1968,Ruffini:1969qy,Liebling_2023}. 
Binary systems of two BSs have already received significant attention as potential GW sources \cite{Palenzuela_2017,Bezares_2022, Helfer_2022, Sanchis_Gual_2022_proca, Evstafyeva_2023, Siemonsen_2023, Ge:2024itl}. 
Even for moderate compactness, the GW emission of BS binaries has been shown to exhibit significant degeneracies with that from standard BH binaries \cite{Evstafyeva:2024qvp}, although deviations from binary-black-hole (BBH) waveforms
may be identified through inference with advanced BBH approximants and elaborate tests \cite{Pompili:2025cdc}.
 BSs have furthermore received attention due to their ability to mimic a wide range of BH phenomenology \cite{Torres:2000dw, Guzman:2005bs, Amaro_Seoane_2010,Rosa_2022sup, Rosa_2022_mim, Rosa_2023}, and as candidate objects for dark matter haloes \cite{Sin_1994, Schive_2014}.

There are, however, new challenges presented by binaries of BSs and related objects when we turn to the creation of waveform template banks. 
This is largely due to the importance of the field interaction. Relative offsets in the BS phase have been found to significantly affect observed GW signals, potentially introducing smoking-gun effects \cite{Palenzuela:2006wp,Palenzuela:2007dm,Evstafyeva:2022bpr,Evstafyeva:2024qvp}. 
This introduces additional free parameters that must be accounted for in any bank of high-accuracy waveforms; see e.g.~\cite{Damour_2025}.
Furthermore, there are a large number of field models to choose from. 
Even restricting ourselves to BSs consisting of a single complex scalar field, a large number of potentials have been proposed (see Ref.~\cite{Liebling_2023} for a review), often introducing still more free parameters, and this is to say nothing of alternatives such as Proca (spin-1) stars \cite{Brito_2016}
or $\ell$-BSs \cite{Alcubierre:2018ahf}.
Because of this high degree of complexity, it is desirable to  start the construction of template banks by considering simplified scenarios in which the relevant number of parameters is smaller.

Such a scenario is provided by binary mergers involving one classical BH and one ECO. 
An obvious simplifying feature of this type of event arises from the absence of a physically relevant scalar-phase offset, assuming the BH is hairless. 
A further simplification relative to BS binaries is suggested by drawing an analogy with the case of BH-neutron star (NS) binaries: there the GW signal depends primarily on the mass ratio and compactness of the NS, with the exact equation of state having only a minor effect \cite{Duez:2009yy, Kyutoku_2010}. 
If true also for BSs, this effect may greatly simplify the generation of a BH-BS or similar template bank, applicable across different classes of ECOs.
Of course, if there is only a small number of ECOs among a larger BH population, we would expect BH-ECO mergers to be more common than ECO-ECO purely on statistical grounds.

So far, however, BH-ECO mergers have received relatively little attention in the literature compared to their ECO-ECO and, in particular, BS-BS counterparts. 
References \cite{Cardoso_2022,Zhong_2023} present results for head-on collisions of a BH and BS, focusing particularly on the case in which the BS mass is significantly larger than that of the BH, and finding that the scalar matter is nonetheless almost entirely accreted.
BH-BS collisions with very light BSs have also been performed as part of a ringdown study \cite{Siemonsen:2024snb}.
Reference~\cite{Clough2018axion} presents a series of head-on collisions between axion stars (ASs) and BHs/NSs: in the case of AS-BH collisions, they find that less compact stars and spinning BHs can decrease the amount of scalar matter that is promptly accreted during the merger. 
BH-BS mergers have also been simulated as part of an exploration of BS-BS collisions involving unstable stars; here one or both BSs may pre-collapse to BHs prior to merger \cite{Ge:2024itl}.
An alternative scenario of reduced complexity concerns the so-called ``dark'' BS binaries, where the BSs couple  exclusively through their gravitational interaction \cite{Bezares_2018}.
Finally, there has been work done on the phenomenological impact of scalar clouds surrounding BH-BH binaries, using both analytical \cite{Tomaselli:2024dbw, Boskovic:2025ixx} and numerical \cite{Aurrekoetxea_2024} tools. 
{Soon after publication of our results, a similar study of BH-BS head-on collisions~\cite{Ning:2026qxs} appeared, paying particular attention to higher multipoles of the GW signal, and finding results compatible with ours.}

In this work, we explore the dynamics of BS-BH mergers, focusing on the impact of varying the scalar potential while holding the BS mass and compactness constant. 
Bearing in mind that the compactness of a BS is not unambiguously defined, our study also addresses the question: \textit{which definition of compactness most systematically
captures GW signatures?}

Our paper is structured as follows.
In Section~\ref{sec:theory}, we briefly review the BS models under consideration and lay out the numerical framework governing our evolutions. 
In Section~\ref{sec:initial_data}, we discuss methods for constructing accurate initial data for BS-BH binaries. 
The bulk of our results is contained in Section~\ref{sec:headon}, where we present a large series of simulated head-on BS-BH collisions.
We determine the dependence of the GW emission on the scalar potential across a broad range of compactnesses and initial boost velocities.
We also present runs colliding ultracompact BS models with BHs, for which we perform additional analysis.
Finally, in Section~\ref{sec:inspirals} we present some preliminary results for BS-BH inspirals, which, to the best of our knowledge, are new to the literature. These suggest that, contrary to the NS case, the tidal disruption of BSs by BHs depends significantly on details of the matter model, and may serve as a formation mechanism for rotating gravitational-atom configurations of a BH surrounded by a co-rotating scalar cloud \cite{Baumann_2019}. 
This is particularly interesting in light of recent numerical work suggesting that such configurations can be dynamically stable on long timescales \cite{Carretero_2025}, at least for configurations for which the mass stored in the scalar matter does not dominate the BH mass \cite{Nicoules_2025}.

Throughout, we work in natural units setting $\hbar = c = G = 1.$  We label spacetime indices by Greek letters running from 0 to 3 and spatial indices by Latin letters running from 1 to 3.

%=============================================================================
\section{Theory and Numerical Methodology}
\label{sec:theory}
\subsection{Review of Boson Stars}
\label{sec:review_BS}
We consider general relativity minimally coupled to a complex scalar field $\varphi$ with 
potential $V(|\varphi|^2)$ via the action,
\begin{equation}
    \!\!\!
    \mathcal{S} = \int \du ^4x\frac{\sqrt{-g}}{2}\left\{\frac{R}{8 \pi}  -\left[
  g^{\mu\nu}\nabla_{\mu}\bar{\varphi}\,\nabla_{\nu}\varphi
  +V(|\varphi|^2)\right] \right\}\,. \label{eq:action}
\end{equation}
Here $R$ is the Ricci scalar, $g_{\mu\nu}$ is the spacetime metric, and an overbar denotes complex conjugation. Varying $\mathcal{S}$, we obtain the Einstein-Klein-Gordon equations:
\begin{eqnarray}
  && R_{\alpha\beta} - \frac{1}{2}R g_{\mu\nu}=8\pi \,T_{\alpha\beta}\,,
  ~~~~~
  \nabla^{\mu}\nabla_{\mu}\varphi = \frac{\du V}{\du \bar{\varphi}}\,, \label{eq:EKG}
  \\[5pt]
  && T_{\mu\nu}
  = \nabla_{(\mu} \bar{\varphi}\,\nabla_{\nu)}\varphi
  -\frac{1}{2}g_{\mu\nu}
  \left[
  g^{\alpha\beta}\nabla_{\alpha}\bar{\varphi}\,\nabla_{\beta}\varphi
  + V(\varphi)
  \right].
  \nonumber
\end{eqnarray}
In this work we restrict our attention to binaries composed of spherically symmetric BSs. Such BS models are obtained by using a harmonic ansatz $\varphi = A(r)e^{\iu\omega t}$ for the scalar field and a line
element given in polar-areal coordinates by
\begin{equation}
  \du s^2 = -\alpha(r)^2\du t^2 + X(r)^2\du r^2 + r^2\du \Omega^2.
\end{equation}
The values of the BS frequency $\omega$ that result in asymptotically flat spacetimes form a discrete ordered set. Here, we consider only BSs in the ground state, so that the field profile $A(r)$ has no zero crossings.
 We compute these models by solving for $\omega$ using a shooting method as detailed, for example, in Ref.~\cite{Evstafyeva:2023kfg}.

We will employ models with two
choices of scalar-potential families. 
The first adds a quartic self-interaction term to the free Klein-Gordon potential,
\begin{equation}
    V_\mathrm{massive}(|\varphi|^2) = \mu^2 |\varphi|^2 + \frac{\lambda}{2}|\varphi|^4
\end{equation}
for $\lambda$ constant.
Models with $\lambda = 0$ are referred to as \textit{mini} BSs, while those with $\lambda > 0$, corresponding to a repulsive self-interaction, are often referred to as \textit{massive} BSs.  
The presence of this repulsive interaction raises the maximal 
BS mass $\mu M_{\rm BS}$ and compactness achievable in the perturbatively stable regime relative to the mini BS case; see e.g. Ref.~\cite{Kain:2021rmk}.
Generally, we will present results in terms of the dimensionless parameter $\hat\lambda \defeq \lambda / \mu^2$.

We also consider \textit{solitonic} potentials of the form
\begin{equation} \label{eq:solitonic}
    V_\mathrm{solitonic}(|\varphi|^2)= \mu^2 |\varphi|^2 \left(
  1-2\frac{|\varphi|^2}{\sigma_0^2}
  \right)^2,
\end{equation}
which are characterized by the parameter $\sigma_0 > 0$. 
Notice that the solitonic potential supports degenerate vacuum states, at $|\varphi| = 0$ and additionally at $|\varphi| = \sigma_0 / \sqrt{2}$. 
This allows for BS models whose scalar-field
amplitude resides close to the latter ``false'' vacuum state
in the stellar interior and transitions to the
$\varphi=0$ vacuum within a narrow \textit{thin-shell} range. These (nonrotating) BSs can achieve very high compactness and mass, even to the extent of possessing light rings, despite 
the fact that the {\it leading-order} correction to the Klein-Gordon potential is attractive. {Furthermore, different scalar potentials are known to result in very different tidal properties \cite{Sennett_2017}. We comment on this further in Sec.~\ref{sec:potential_impact}.}

The $U(1)$ symmetry of the action \eqref{eq:action} is associated with a conserved current $J^\mu = \frac{\iu}{2}(\varphi\nabla^\mu\bar\varphi - \bar\varphi\nabla^\mu \varphi).$ 
This gives rise to a conserved Noether charge, which one can interpret as a particle number,
\begin{equation} \label{eq:noether}
    N = \int \du ^ 3 x \;\sqrt{-g} \;J^0 .
\end{equation}

Notice that for a chosen potential and with the harmonic ansatz $\varphi=A(r)e^{\iu \omega t},$  our action \eqref{eq:action} is invariant under the rescaling,
\begin{align}
    \mu &\rightarrow a\mu, \; x^\nu \rightarrow \frac{x^\nu}{a}, \; \lambda \rightarrow a^2\mu, \; \omega \rightarrow a\omega
\end{align}
for any constant $a > 0$, with the BS mass and Noether charge scaling as $M_\mathrm{BS} \rightarrow aM_\mathrm{BS}$
and $N \rightarrow a^2 N$ respectively.

Henceforth, we will always choose $a$ such that $M_{\rm BS}+M_{\rm BH}=1$, i.e.~the BS mass depends only on the mass ratio of our binary, which simplifies the construction of our computational grid in code units. We also define the \textit{binding energy} $E_B \defeq M_\mathrm{BS} - \mu N$, measuring the difference in energy between the stationary BS configuration and a situation in which all the scalar matter is at rest at infinity.
In this convention, BSs with negative binding energy are interpreted as gravitationally bound, and those with positive binding energy are gravitationally unbound.

\begin{figure}
    \includegraphics[width=\linewidth]{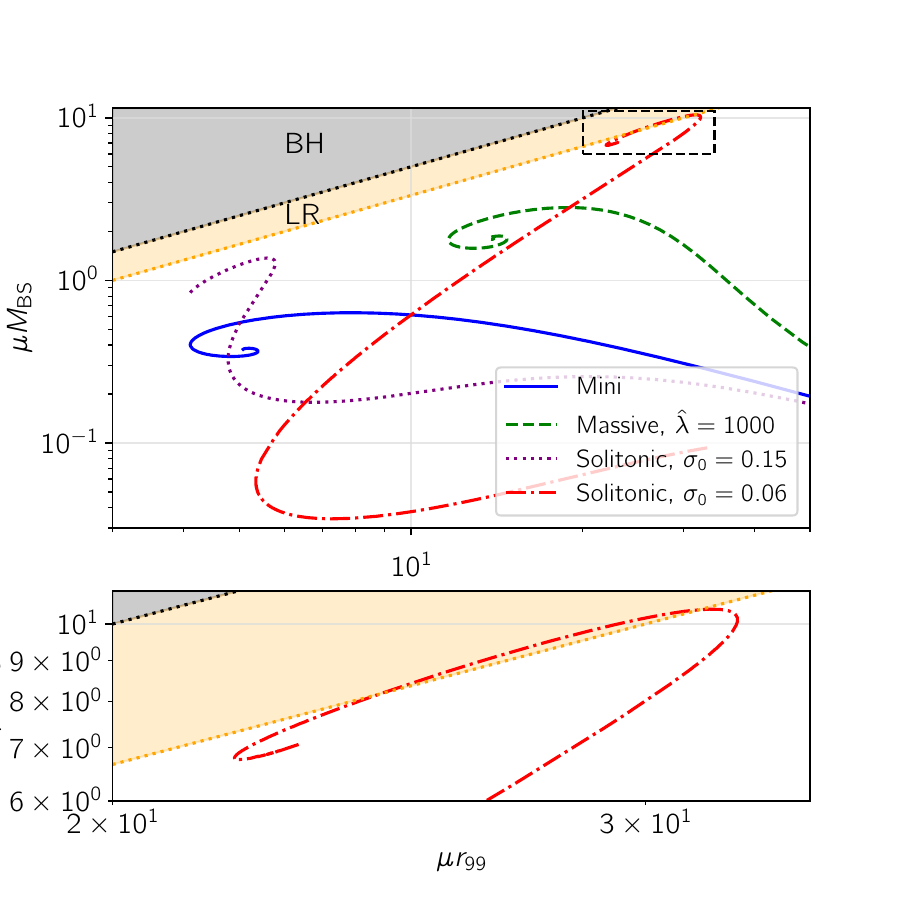}
    \caption{
    Mass $M_\mathrm{BS}$ against radius $r_{99}$ for a selection of spherically symmetric mini, massive, and solitonic BS families. We also shade the regions $M_\mathrm{BS} / r_{99} > 1/2$ and $M_\mathrm{BS} / r_{99} > 1/3$, which correspond to the interior of the event horizon and light ring, respectively, for a Schwarzschild BH. The bottom panel zooms in on where the curve for solitonic BSs with $\sigma_0 = 0.06$ intersects this latter region; these models are indeed ultracompact with a pair of light rings.
    }
    \label{fig:BS_families}
\end{figure}

Because of the exponential decay of the scalar field profile, BSs do not have a hard surface. 
We therefore adopt the common convention of defining the (areal) BS radius $r_{99}$ as that enclosing 99\% of the mass. 
In Fig.~\ref{fig:BS_families}, we plot the mass against this radius for a selection of mini, massive, and solitonic families that we will use frequently in this work, including one family that includes perturbatively stable UCOs.

Finally, we briefly discuss the topic of linear BS stability. 
Much like Tolman-Oppenheimer-Volkoff \cite{Tolman:1939jz,Oppenheimer:1939ne} stars, the parameter space for a given BS family is split into stable and unstable branches, separated by local extrema in the mass $M_\mathrm{BS}$ (or, equivalently, Noether charge $N$) \cite{Gleiser:1988rq}.
Such extrema generically correspond to zero-frequency modes in the BS's radial-oscillation spectrum \cite{Gleiser:1988ih,Santos:2024vdm}.
The existence of these modes therefore provides a necessary, albeit not sufficient, condition for a transition between stable and unstable branches of a BS family.
 More specifically, mini and massive BSs feature only one stable branch,
ranging from the vacuum limit $A_0\to 0$ (where $A_0$ denotes the central scalar field amplitude) to the BS model of (globally) maximal mass \cite{Kain:2021rmk}.
Solitonic potentials with sufficiently small $\sigma_0$, however, admit an additional second stable branch connecting the first local minimum in the function $M_{\rm BS}(A_0)$ to the second maximum \cite{Santos:2024vdm,Marks_2025}.

In this work, we focus exclusively on binaries composed of stable
BSs, but we note in passing that unstable BS models can play
important roles in theoretical investigations, as for example
by serving as the critical solutions in the collapse of massive
complex scalar fields \cite{Hawley_2000, Jimenez_2022}. Our focus on stable BSs
naturally implies some restrictions on the range of compactness available for a given potential;
cf. Fig.~11 in Ref.~\cite{Marks_2025_B}.
In particular, all the solitonic models we consider in this work are found on the second (higher compactness) stable branch; for the $\sigma_0$ values considered here, the stars on the first stable branch are of rather low compactness and
correspondingly inefficient in generating GWs.

\subsection{Compactness Definitions} \label {sec:compactness_definitions}
As  noted earlier, BSs lack a hard surface, so that their compactness cannot be unambiguously defined. Here, we will briefly discuss several alternative compactness definitions  considered in this work. Their suitability as representative parameters for capturing GW signatures will be investigated in Sec.~\ref{sec:potential_impact}.

A common convention is to define the BS compactness in terms of the above-mentioned areal radius $r_{99}$ enclosing 99\% of the aspect mass,
\begin{equation}
    C_0 \defeq \frac{M_\mathrm{BS}}{r_{99}}.
\end{equation}
The suitability of various definitions of compactness has been discussed in Refs.~\cite{Collodel:2022jly, Boskovic:2021nfs}.
In the former, the authors compared five plausible definitions for solitonic BSs including UCOs.
They identified that three of these definitions exhibited good agreement, especially for more compact models, and were furthermore free of pathologies such as the compactness exceeding $1/3$ before a pair of light rings emerged.
We will therefore also consider one of their ``good'' definitions,  namely,
\begin{equation}
C_4 \defeq \frac{M_\mathrm{BS}} {r_4},
\end{equation}  where $M_\mathrm{BS}$ is the BS mass and $r_4$ is the radius at which the radial component of the BS metric in polar-areal coordinates agrees with the Schwarzschild value to 99.9\% accuracy. 

Both $C_0$ and $C_4$ depend strongly on the BS profile at large radii.
There is reason, however, to believe that the distribution of scalar matter at large radii may be significantly less important for understanding their dynamics than that in the bulk.
For instance, Ref.~\cite{Hawley_2000} studies the critical collapse of mini BS solutions perturbed by a massless real scalar field, finding that the critical solution is well-described by a BS on the unstable branch.
Due to the formation of ``haloes" of scalar matter, the critical solutions depart significantly from equilibrium BSs at large radii, yet nonetheless agree very well with the predictions made by applying linear perturbation theory to an (unstable) equilibrium configuration.
 In view of these findings, we also consider a definition of the compactness designed to predominantly capture the distribution of scalar matter in the bulk rather than at larger radii,
\begin{equation} \label{eq:c_max}
    C_\mathrm{max} \defeq \max_{r > 0} \frac{m(r)}{r},
\end{equation}
where $m$ is the aspect mass enclosed within a radius $r$.

%==================================

\subsection{Computational Framework}
\label{sec:computational_framework}
The BS-BH mergers presented in this work have been simulated using {\sc grchombo} \cite{Andrade:2021rbd, Radia:2021smk}, an open-source code capable of evolving the the Baumgarte-Shapiro-Shibata-Nakamura (BSSN) \cite{Baumgarte:1998te,Shibata:1995we} and conformal covariant Z4 (CCZ4) \cite{Alic:2011gg} formulations of numerical relativity with full adaptive mesh refinement (AMR). 
In this work, we exclusively use the CCZ4 formulation due to its constraint-damping properties; cf.~Refs.~\cite{Evstafyeva:2025mvx, Marks_2025_B}, in which dynamical evolutions of isolated, stationary BSs are seen to exhibit long-lasting linear
growth in the Hamiltonian constraint violation when evolved with BSSN but not CCZ4.
The full evolution equations used in this study are given in Appendix~\ref{sec:evolution_equations}.
For the damping parameters, we set $\kappa_1 = 0.1 / (M\alpha)$, $\kappa_2 = 0, \kappa_3 = 1,$ where henceforth $M$  denotes the Arnowit-Deser-Misner (ADM) \cite{Arnowitt:1962hi} mass of the spacetime and $\alpha$ is the lapse. 
In addition, we make use of the {\sc exozvezda} extension, whose capabilities are outlined in Refs.~\cite{Helfer_2022,Evstafyeva:2023kfg,Croft:2022bxq, Croft:2022gks}, to dynamically evolve scalar matter and compute fluxes. 

For the head-on collisions studied in this paper, we make use of a dimensional reduction of {\sc grchombo} to enforce axisymmetry by means of the so-called ``modified cartoon'' method \cite{Pretorius_2004, Yoshino:2009xp, Cook:2016soy}.
Below we will often refer to these simulations as ``2+1'' but
emphasize that they model full 3+1 spacetimes merely using an
effective two-dimensional domain.

Unless otherwise indicated, each run presented in this work has been performed on a grid of edge length $L =1024 M$ 
with 7 AMR levels present. The grid spacing on the finest level is $\Delta x = M/32$ for inspirals and $\Delta x = M/64$ for head-on collisions, where the higher resolution allows for the larger boost velocities
$v\lesssim 0.7$ used in Sec.~\ref{sec:boosts} below.
Our tagging criterion is based on second derivatives of the conformal factor, scalar field, and energy density; see Section 3.5 of Ref.~\cite{Radia:2021smk} for more details.
We use a radius $R_{\rm ex} = 140 M$ for gravitational-wave extraction.
A more detailed error analysis, including convergence tests and error budgeting due to truncation and extraction at finite radius, is included in Appendix~\ref{app: convergence and error analysis}. 
The main conclusion is that our reported GW energies are accurate to within $\sim1.5\%$ for our default head-on collision of equal-mass collisions with boost velocity $v = 0.1$, remain accurate to within $\sim3\%$ for the most extreme boost velocities and mass ratios considered, while the error is $\sim 6\,\%$ for our inspirals.

%=============================================================================
\section{Initial Data}
\label{sec:initial_data}

In this section, we review the construction of initial data for binaries involving BSs, and discuss methods particular to the case of BS-BH binaries. 
We label the  
 3+1 spatial metric and extrinsic curvature for an isolated BH and BS in equilibrium by $\{\gamma_{ij}^\mathrm{BH}, K_{ij}^\mathrm{BH} \} $ and $\{\gamma_{ij}^\mathrm{BS}, K_{ij}^\mathrm{BS} \} $, respectively. 
In this work we restrict our attention to spherically symmetric BHs and BSs in isotropic coordinates, but the considerations here would not change significantly should one wish to construct binaries involving arbitrary Bowen-York BHs or rotating BSs instead.

The most straightforward approach to constructing binary initial data, which we will call \textit{plain superposition}, involves simply adding the metric and extrinsic curvature tensors such that asymptotic flatness is preserved,
\begin{align}
\gamma_{ij} &= \gamma_{ij}^{\mathrm{BH}} + \gamma_{ij}^{\mathrm{BS}} - \delta_{ij} ,\\
K_{ij} &= \gamma_{m(} [K^{\mathrm{BH}}_{j)n} \gamma_\mathrm{BH}^{nm} + K^{\mathrm{BS}}_{j)n} \gamma_\mathrm{BS}^{nm} ].
\end{align}
Plain superposition has been used in prior studies of head-on BS-BH collisions \cite{Cardoso_2022, Zhong_2023}, where its use was justified in part by the fact that the BH mass was significantly smaller than that of the BS, reducing the impact of the perturbation to the BS. 
In practice, however, this procedure has been found to be problematic in the modeling of comparable-mass BS-BS binaries \cite{Helfer_2022}, mainly for two reasons:  (i) Plain superposition
introduces significant constraint violations, and (ii) the volume element in the BS interior is changed relative to the equilibrium configuration while the scalar field profile remains the same, effectively preparing the BS in a perturbed state.
The resulting radial oscillations of the BSs are often long-lived \cite{Seidel_Suen_1990, Lai_2007, Marks_2025}, thereby remain relevant throughout the inspiral process, and can introduce unphysical artefacts into the extracted GW signal.
For compact BSs close to a perturbatively unstable branch, this effect can drive an unphysical pre-merger collapse into a BH.

The simple remedy proposed in Ref.~\cite{Helfer_2022} can be straightforwardly adapted to the case of BS-BH binaries.
One only has to prepare the spatial metric using the modified superposition,
\begin{equation}
    \gamma_{ij}(x^k) = \gamma_{ij}^{\mathrm{BH}}(x^k) + \gamma_{ij}^{\mathrm{BS}}(x^k) - \gamma_{ij}^{\mathrm{BH}}(x^k_\mathrm{BS}) \label{eq:metric_correction}
\end{equation}
where $x^k_\mathrm{BS}$ denotes the coordinates of the BS center.
By restoring the volume element corresponding to an equilibrium configuration at the BS center, this greatly reduces the spurious oscillations that otherwise plague BS evolutions. 
We will refer to initial data prepared according to this method as \textit{metric-corrected}.

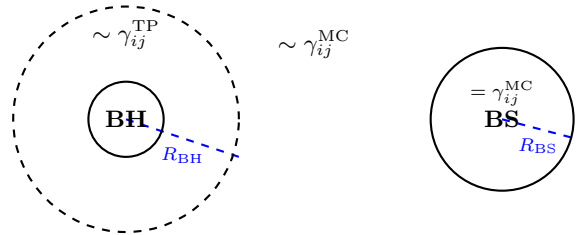
\begin{figure}[t]
    \centering
    \vspace{10pt}
    \begin{tikzpicture}[>=latex, font=\small]
% Parameters
\def\BHradius{0.5}       % Inner (solid) circle radius for BH
\def\BHouter{1.5}        % Outer (dashed) circle radius for BH
\def\BSradius{0.95}       % Solid circle radius for BS
\def\sep{5.0}            % Horizontal separation between BH and BS

% --- BH structure ---
\coordinate (BHcenter) at (0,0);

% Line and label (R_BH)
\draw[thick, dashed, blue] (BHcenter) -- ++(\BHouter,-0.5)
    node[midway, below, font=\scriptsize] {$R_{\mathrm{BH}}$};

\draw[thick] (BHcenter) circle (\BHradius);
\node at (BHcenter) {\textbf{BH}};
\draw[dashed, thick] (BHcenter) circle (\BHouter);

% Text between circles (BH)
\node at ($(BHcenter)+(0,0.75*\BHouter)$) {$\sim\gamma_{ij}^\mathrm{TP}$};

% --- BS structure ---

% Line and label (R_BS)
\coordinate (BScenter) at (\sep,0);
\draw[thick, dashed, blue] (BScenter) -- ++(\BSradius,-0.25)
    node[midway, below, font=\scriptsize] {$R_{\mathrm{BS}}$};
    
\draw[thick] (BScenter) circle (\BSradius);
\node at (BScenter) {\textbf{BS}};
\node[above=2pt of BScenter, font=\scriptsize] {$=\gamma_{ij}^\mathrm{MC}$};

% --- Text between BH and BS ---
\node at ($0.5*(BHcenter)+0.5*(BScenter)+(0,1.0)$) {$\sim\gamma_{ij}^\mathrm{MC}$};

\end{tikzpicture}
    \caption{Schematic representation of the correction to the spatial metric constructed using {\sc twopunctures}-based initial data for an  analogous BH-BH binary: (i) in a region surrounding the BH, the data are close to those computed by {\sc twopunctures}; (ii) in a region surrounding the BS, the data are given by the metric-correction~\eqref{eq:metric_correction}; (iii) everywhere else 
    the data are the metric-corrected version slightly modified by the attenuation function $f(x^i)$
    of Eq.~(\ref{eq:fofx}). The interpolating region around a radius $R_\mathrm{BH}$ surrounding the BH inevitably introduces some residual constraint violation; this region is intentionally kept away from the BS to avoid exciting radial oscillations as much as possible.}
    \label{fig:twopunctures_schematic}
\end{figure}

While these initial data are not fully constraint solved, the constraint violations surrounding the BS are significantly reduced relative to plain superposition.
By time evolving BS-binary initial data constructed
with metric correction (termed ``constant-volume-element data'' in their paper) with and without subsequent elliptic solving of the
constraint equations, Ref.~\cite{Atteneder_2024} found the metric correction to be the dominant ingredient for improving the
extracted GW signals and other physical diagnostics.
Likely, the constraint damping of the CCZ4 formalism further reduces the adverse impact of the initial data's residual constraint violations.

\begin{figure}
    \includegraphics[width=\linewidth]{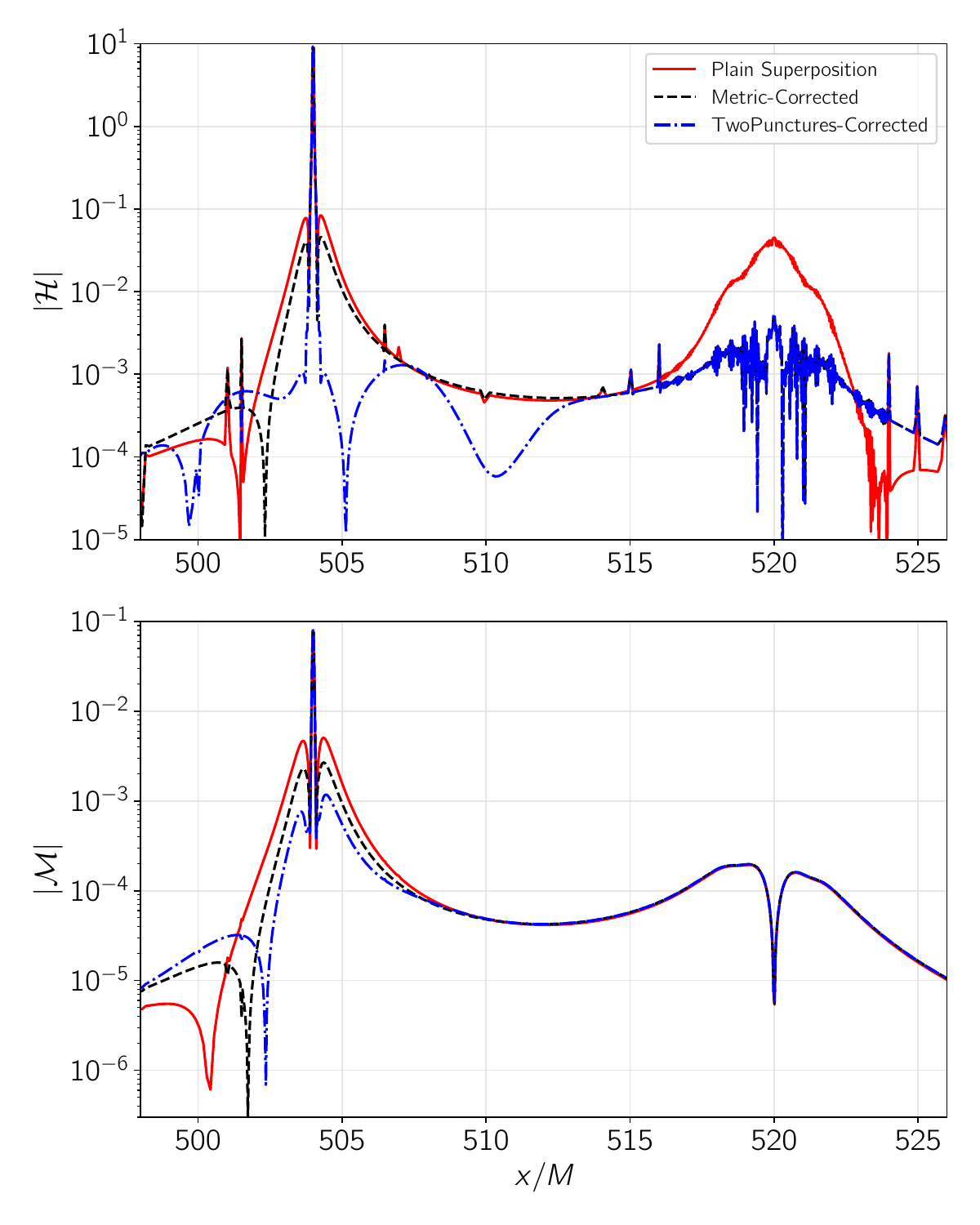}
    \caption{
    The Hamiltonian (top) and
    magnitude
    $\mathcal{M}=\sqrt{\mathcal{M}_x^2+\mathcal{M}_y^2+\mathcal{M}_z^2}$
    of the momentum
    (bottom) constraint violations
    for the initial data of an equal-mass BH-BS head-on collision with initial separation $d=16M$, boost velocity
    $v=0.1$ in the center-of-momentum frame, and a solitonic BS
    (on the right) with $\sigma_0=0.2$ and central scalar amplitude
    $A_0=0.17$. The three curves show the constraints
    for
    plain superposition, metric- and {\sc TwoPunctures}-corrected initial data obtained for $R_\mathrm{BH} = R_\mathrm{BS} = 4M.$ The large spike at $x\approx 504M$ present in both plots corresponds to the interior of the BH's event horizon and hence
    %does not propagate outwards To cover against superluminal gauge speeds, perhaps rephrase to ``
    remains causally disconnected from the exterior spacetime. The smaller spikes are merely diagnostic artefacts due to the finite differencing at mesh refinement boundaries
    and do not affect the time evolution.
    }
    \label{fig:constraint_lineouts}
\end{figure}

After metric-correcting our data, the remaining constraint violations are mainly located close to the BH. 
A simple approach to reducing these, without disturbing the BS from equilibrium, is to exploit Birkhoff's theorem and the exponential decay of the scalar field. Specifically, we can replace in a region around the BH the metric corrected data with fully constraint-solved initial data for a BH-BH binary in which the second BH has mass and momentum equal to that of the BS. 
To this end, we make use of the {\sc twopunctures} initial data solver \cite{Ansorg:2004ds} for an appropriate BH-BH binary, labeling the resulting initial data  $\{\gamma_{ij}^\mathrm{TP}, K_{ij}^\mathrm{TP} \} $
We then modify the spatial metric correction according to the prescription,
\begin{align}
\gamma_{ij}^\mathrm{MC}(x^k) &\defeq \gamma_{ij}^{\mathrm{BH}}(x^k) + \gamma_{ij}^{\mathrm{BS}}(x^k)  - \gamma_{ij}^{\mathrm{BH}}(x^k_\mathrm{BS}), \\
\label{eq:twopunctures_correction }\gamma_{ij}(x^k)   = \, & \gamma_{ij}^\mathrm{MC}(x^k)  & \\ \nonumber  &+ f(x^i) \left[ \gamma_{ij}^\mathrm{TP}(x^k) - \gamma_{ij}^\mathrm{MC}(x^k) \right],
\end{align}
where $f$ is a smooth function satisfying $f(x_\mathrm{BH}^k) = 1$ and $f \rightarrow 0$ at spatial infinity. 
Defining $r_\mathrm{BS}$ and $r_\mathrm{BH}$ as the coordinate distances to the BS and BH center, respectively, we empirically found an effective choice to be given by
\begin{equation}
    f(x^i) = H(r_\mathrm{BS} - R_\mathrm{BS})\left\{1 - \tanh\left[\left(r_{\rm BH} / R_{\mathrm{BH}}\right)^2\right]\right\}\,,
    \label{eq:fofx}
\end{equation}
where $R_{\mathrm{BH}}$ and $R_{\mathrm{BS}}$ are constants and $H$ is the Heaviside step function, included to ensure the initial data near the BS are not influenced by the coordinate singularity associated with the second BH used by {\sc twopunctures}. The analogous correction is also made for the extrinsic curvature $K_{ij}$.  We illustrate the idea behind this correction schematically in Fig.~\ref{fig:twopunctures_schematic}.

In general, we find choosing $R_{\mathrm{BH}}$ 5 to 10 times the BH mass to be optimal for minimizing constraint violations, while $R_{\mathrm{BS}}$ can be simply taken to be the BS radius $r_{99}$ defined in Sec.~\ref{sec:review_BS}. 
In Fig.~\ref{fig:constraint_lineouts}, we illustrate the
improvements thus obtained for an initial snapshot
of a representative BH-BS head-on collision.

Ideally the above metric correction should be succeeded by
an elliptic solving of all four constraint equations.
From a physical viewpoint, however, this faces a key challenge: preparing equilibrated data involves fixing profiles for the matter and conformal factor $\psi = \left[\det\gamma\right]^{1/12}$ simultaneously, while typical approaches to elliptic solving of the constraint equations, including the conformal thin-sandwich method used in Refs.~\cite{Atteneder_2024, Siemonsen_2023}, solve for $\psi$ with fixed matter variables, thus typically disturbing the equilibrium.
This challenge is addressed in Ref.~\cite{Siemonsen_2023} which provides methods for reducing the impact of this disturbance on GW signals, and suggests that they might be removed entirely by solving quasi-equilibrium equations for the scalar matter simultaneously with the constraint equations.

As an alternative approach, one may use the conformal-transverse-traceless-K (CTTK) formalism~\cite{Aurrekoetxea_2023} to prescribe arbitrary profiles for the conformal factor by converting the Hamiltonian constraint into an algebraic equation for the trace of the extrinsic curvature $K$.
At present, however, the CTTK formalism has been developed and tested only in the case of conformally flat initial data,  where $\gamma_{ij} = \psi^4\delta_{ij}$.
For nonzero initial velocities, this condition is not met by the ``initial guess'' constructed with the superposition methods discussed above, so that either the
CTTK method or our superposed initial guesses need adjustment.
We leave an investigation of possible resolutions of this
conundrum for future work. {We also highlight the recent alternative approaches to constructing initial data proposed by Refs.~\cite{Palloni:2026phz, Doherty:2026yvb}, which appeared shortly after the original version of this paper.}

%=============================================================================

%
%=============================================================================
\section{Head-On Collisions}
\label{sec:headon}
In this section, we present results for BS-BH head-on collisions, dynamically evolved in enforced axisymmetry. 
Unless otherwise indicated, all collisions are performed with an initial separation of $d = 80M$. 

\subsection{Plain Superposition vs. Metric Correction}
\label{sec:plain_vs_metric}

To underscore the importance of preparing the BS in an equilibrium configuration, we start by presenting a series of runs comparing plain superposition to metric-corrected initial data.
For this comparison, we consider mini BS models across a wide range of compactness values  $C_{\rm max}$ as defined in Eq.~(\ref{eq:c_max}). The most compact 
of these models is located on the stable solution branch very close to the so-called Kaup limit beyond which all models are perturbatively unstable \cite{Gleiser:1988rq}.

\begin{figure}[t]
    \includegraphics[width=\linewidth]{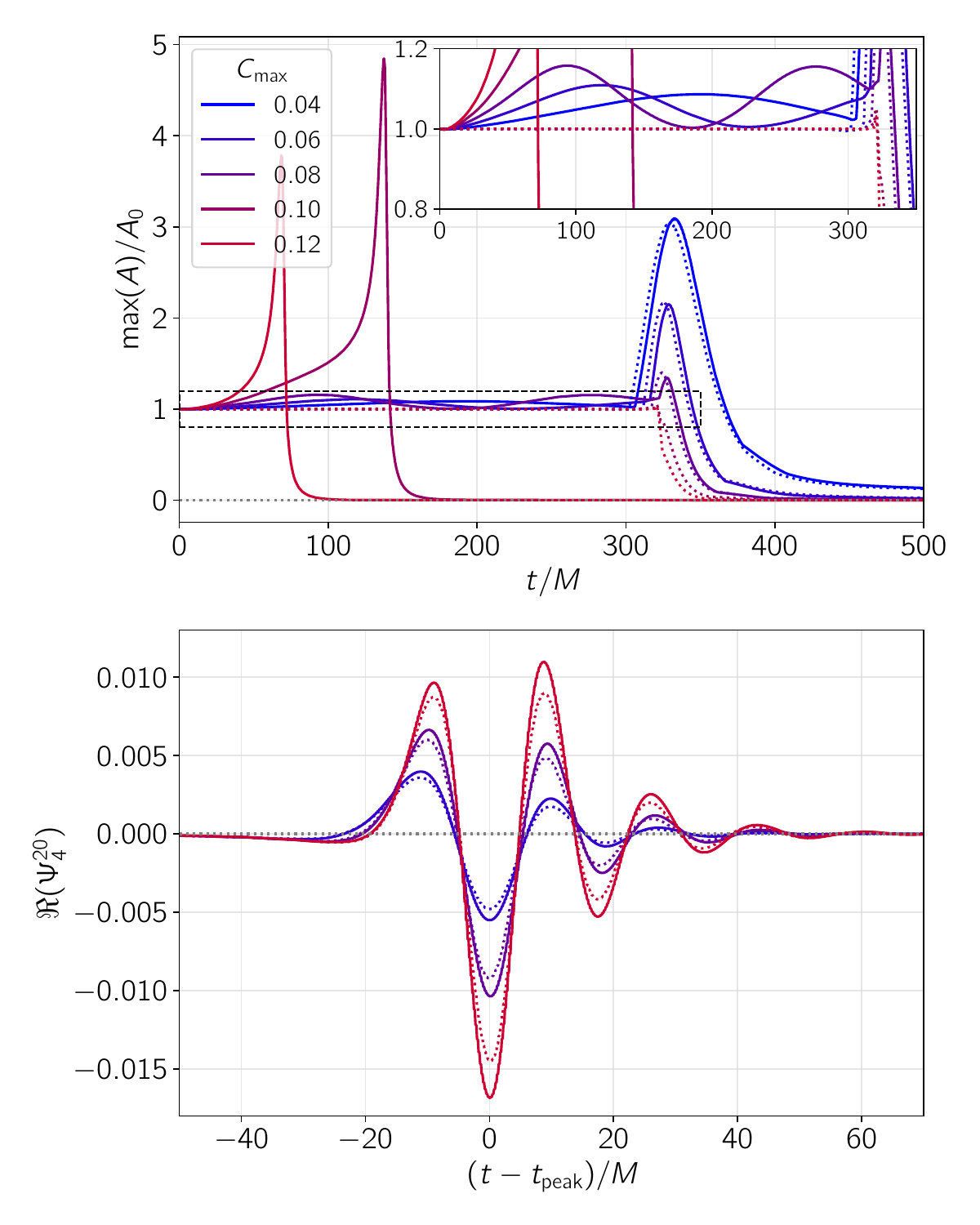}
    \caption{\textit{Top:} Relative change in the maximum scalar-field amplitude
    over time for a series of head-on collisions of one BH and one mini BS
    with various values of the compactness $C_\mathrm{max}$.
    We compare results obtained for initial data constructed with plain superposition (solid) and for metric-corrected initial data (dotted curves).
    Note that models with $C_\mathrm{max} = 0.1$ and $C_\mathrm{max} = 0.12$ collapse into BHs well before merger when prepared using plain superposition.
    The inset shows the behaviour of BSs that do not pre-collapse.
    \textit{Bottom:} The Newman-Penrose scalar $\Psi_4$,
    extracted as detailed in Ref.~\cite{Ge:2024itl},
    for a subset of these head-on collisions. These GW signals have been time-shifted such that the peak amplitudes of all waveforms align at $t=t_{\rm peak}$. They exhibit the
    excess GW radiation effected by the plain superposition.
    }
    \label{fig:plain_vs_metric}
\end{figure}

We display key diagnostics from these runs in Fig.~\ref{fig:plain_vs_metric}.
Even for this relatively large initial separation,  plain superposition significantly impacts the progression of the evolution: for all binary evolutions starting from plain superposition, the BS begins to oscillate at approximately the fundamental radial oscillation frequencies 
computed in Refs.~\cite{Gleiser:1988rq,Hawley_2000}, with the amplitude of these oscillations becoming larger for BSs near the end of the perturbatively stable branch.
Notably, these oscillations are not centered around the central amplitude $A_0$ of the original, stationary BS solution, but instead around a slightly larger value. 
Thus, in addition to supplying kinetic energy through the radial oscillation, the use of plain superposition actually prepares the scalar field in an overdense configuration corresponding to a more massive background BS solution than the one originally considered. 
For models sufficiently close to the unstable branch, there exists no suitable more massive stable background solution, and the BS instead collapses into a BH well before merger. In all cases considered, metric-corrected initial data cure these pathologies.

For the large separations considered in all these
head-on collisions, the impact of the additional constraint
reduction obtained with the \textsc{twopunctures}-based method
is negligible compared to the numerical uncertainties.
In Fig.~\ref{fig:plain_vs_metric}, for example, the corresponding
curves would lie on top of their metric-corrected analogs. In the
remainder of this section, we therefore use metric-corrected
initial data for all head-on collisions.

These considerations help us to explain a phenomenon seen when we turn to the emitted GW signals: runs performed using plain superposition consistently emit more GW energy during merger than those performed using metric-corrected initial data. 
This has been observed previously in the case of BS-BS collisions \cite{Helfer_2022, Evstafyeva_2023}. For models that do not exhibit premature collapse, this is expected on the grounds that the  perturbed BSs resulting from plain superposition have had energy added to both the kinetic and potential sectors.
At first glance, it appears surprising
that this effect is particularly pronounced for models that do exhibit premature collapse.
As we will see in subsequent sections, however, this too is to be expected, since 
equal-mass BS-BH collisions turn out to be generally less radiatively efficient than BH-BH collisions with equivalent masses. In this context,
mixed BS-BH collisions differ from their BS-BS counterparts, which radiate GWs more efficiently across a broad range of compactness \cite{Ge:2024itl}.

\subsection{Impact of the Potential}
\label{sec:potential_impact}

\begin{figure}
    \includegraphics[width=\linewidth]{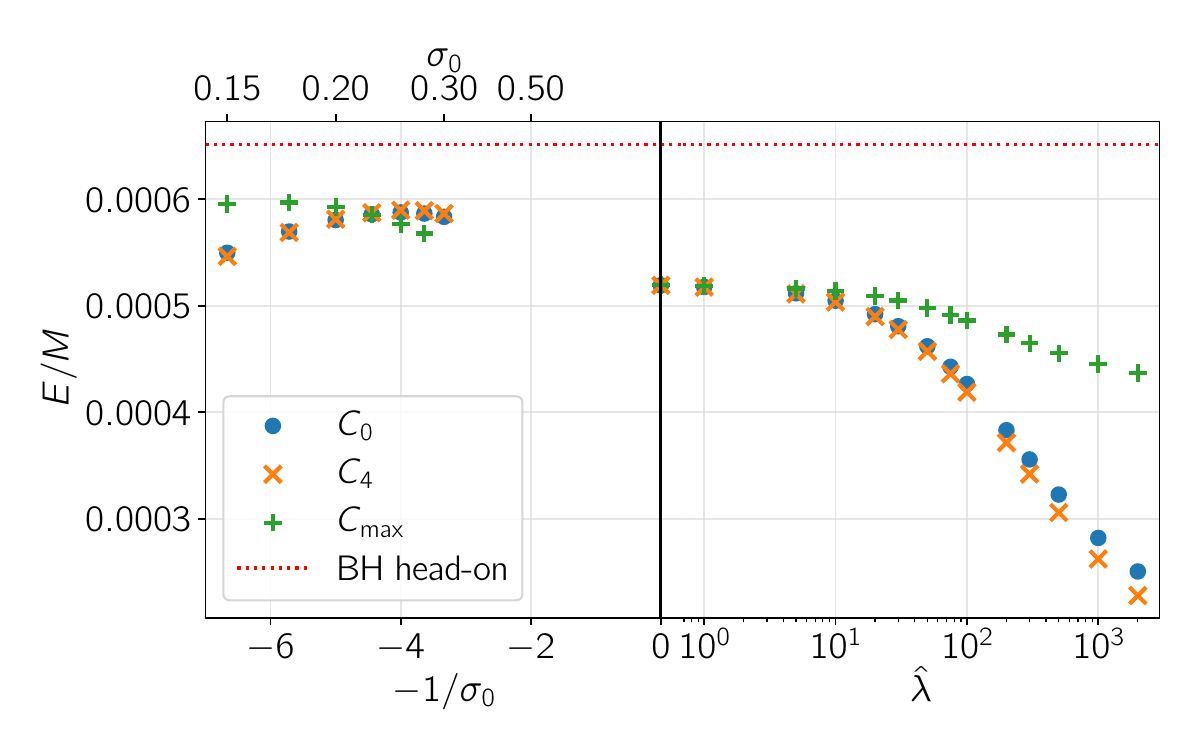}
    \caption{
    Total energy radiated in GWs for a selection of head-on BS-BH collisions for massive and solitonic models with varying $\hat\lambda$ and $\sigma_0$ values. Each BS is chosen such that (a particular definition of) its compactness equals that of a marginally stable mini BS with central amplitude $A_0 = 0.076$. This reference mini BS
    model occupies the shared point at $\hat\lambda = 1/\sigma_0  = 0$. All runs use an initial boost velocity $v = 0.1$ for both objects in the zero-momentum frame. The dotted
    horizontal line marks the GW energy emitted by an equal-mass BH head-on collision with the same boost velocity.
    }
    \label{fig:compactness_definitions}
\end{figure}

\begin{figure*}[ht!]
    \includegraphics[width=\linewidth]{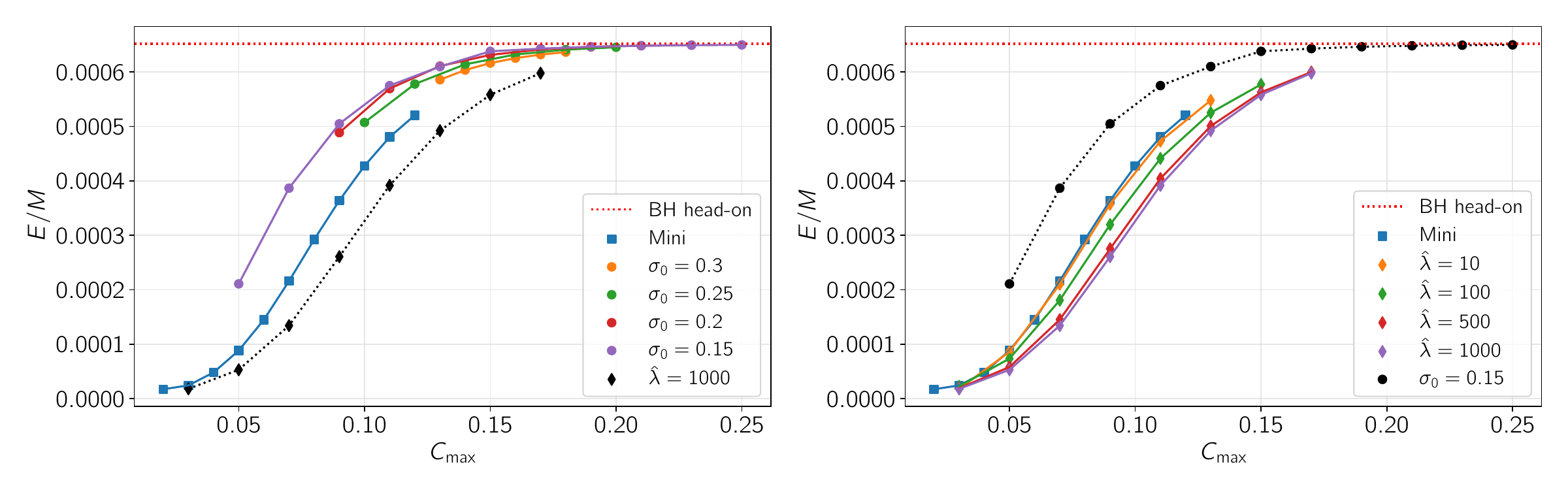}
    \caption{Total energy emitted as GWs for equal-mass head-on BS-BH collisions against maximum compactness $C_\mathrm{max}$, for a range of solitonic (left) and massive (right panel) BS families. The values of $C_\mathrm{max}$ we use span the first stable branch for massive models, and the second for solitonic models. As a visual aid, the most extreme solitonic (massive) family considered is also included in the massive (solitonic) plot.}
    \label{fig:q1_energies}
\end{figure*}

We now turn to a study of how the scalar potential $V(|\varphi|^2)$ affects the phenomenology of BS-BH collisions.
To this end, we will compare runs in which we vary the potential, but tune the central amplitude $A_0$ and scalar mass $\mu$
so that the total mass and compactness (using one of the definitions presented in Sec.~\ref{sec:compactness_definitions}) of the BS remain constant. To assess the influence of these different definitions of compactness, we show in Fig.~\ref{fig:compactness_definitions} the radiated GW energy for a series of equal-mass BS-BH collisions. As our reference case, we here use  a mini BS model with $A_0 = 0.076$, located on the first stable branch and close to the Kaup limit. 
We then perform analogous runs using solitonic and massive BSs, chosen in each family so that (a chosen definition of) the compactness is equal to that of the reference mini BS with compactness parameters\footnote{For comparison, the three compactness measures applied to the ultracompact thin-shell BS model \texttt{S06A044} of Sec.~\ref{sec:bh_uco} are $C_\mathrm{max} \approx 0.335,$ $C_0,  \approx 0.328, C_4\approx 0.322$.}
$C_0 \approx 0.0802$, $ C_4 \approx 0.0744$, $C_\mathrm{max} \approx 0.120$. 

We immediately  see that, irrespective of the definition of compactness used, at fixed compactness the solitonic models radiate more efficiently than the mini model, which is itself more efficient than the massive models. 
All BS-BH collisions emit less energy than the BH-BH equivalent (marked by the horizontal line), a conclusion we will see repeatedly.
Furthermore, the radiative efficiency of massive models at fixed compactness decreases with increasing $\hat\lambda$, again for all definitions of the compactness. 
The extent of this effect is quite striking, particularly when we use $C_0$ or $C_4$ as our definiton: the most extreme massive model considered emits less than half the GW  energy of the most efficient solitonic model, despite having the same mass and ostensibly the same compactness.
This effect is attenuated significantly when we use $C_\mathrm{max}$, suggesting that indeed the distribution of matter in the BS bulk, rather than that in the outer layers, determines the radiative efficiency.
Yet, even for $C_{\rm max}$ the effect is still present.
We can interpret this observation by drawing a loose parallel between the scalar potential and the equation of state (EoS) for a fluid. 
This parallel is well-understood in the case of massive BSs in the limit $\hat\lambda \rightarrow \infty$, for which the scalar matter obeys an effective EoS of the form~\cite{Colpi_1986}
\begin{equation}
    p = \frac 49\rho_0 \left[\sqrt {\left(1 + \frac{3\rho}{4\rho_0}\right)} - 1 \right]^2,
\end{equation}
where $\rho$ and $p$ are the energy density and pressure, respectively, and $\rho_0 \defeq \hat\lambda / (2\mu^2)$ with our conventions. Similar, albeit approximate, correspondences are discussed in Refs.~\cite{Pitz_2023, Faraoni_2012, Chavanis:2025qcg}.
Our observations regarding radiative efficiency suggest we can interpret massive BSs as possessing a ``stiffer" EoS --- compared to mini BSs --- and solitonic BSs a ``softer" EoS\footnote{{This can also be understood by looking at the potential functions $V(|\varphi|)$: for given $|\varphi|$, the massive potential is larger than the mini potential, implying a higher ``cost'' for the scalar field to reach high density, analogous to the effect of larger pressure. Likewise, the negative 4\textsuperscript{th}-order term of the solitonic potential favors large $|\varphi|$, except for highly compact stars with $A_0\gtrsim \sigma_0$ where the 6\textsuperscript{th}-order term in the solitonic potential becomes dominant and the BSs acquire a thin-shell structure.}}. {This analogy helps to further understand the significant differences in compactness and tidal deformability between various scalar potentials.}
This is also consistent with studies of the radial oscillation frequencies of these BS families~\cite{Kain:2021rmk, Marks_2025}.

While massive BSs show a clear decline in radiative efficiency with increasing $\hat\lambda$, the situation is more complicated for solitonic models. 
When using the compactness measure $C_\mathrm{max}$, the radiative efficiency increases with decreasing $\sigma_0$, while this trend is reversed for
$\sigma_0 \lesssim 1/4$ when using $C_0$ or $C_4$ instead.
Furthermore, for both of these, the radiative efficiency varies non-monotonically with $\sigma_0$. 
This is likely due to an increased competition between energy channels in the solitonic case. While the leading-order correction to the Klein-Gordon potential is an attractive interaction, the presence of the degenerate vacuum state at $|\varphi| = \sigma_0/\sqrt{2}$ causes the stars to increasingly approach the so-called \textit{thin-shell} profile, where the energy density is dominated by a sharp peak in the gradient term at finite radius~\cite{Collodel:2022jly, Marks_2025, Boskovic:2021nfs}.
This means that the EoS parallel we can draw for massive models breaks down once we consider sufficiently small $\sigma_0$.

Having established these conclusions for BSs with a compactness close to the stability threshold of mini BSs, we now wish to determine how robust they are across a broader range of compactness values.
From now on, we will focus on $C_\mathrm{max}$ as our definition of the BS compactness.

 We start this discussion by plotting in Fig.~\ref{fig:q1_energies}  the radiated energy against $C_\mathrm{max}$ for a variety of massive and solitonic BS families.
We see that the broad conclusions suggested by Fig.~\ref{fig:compactness_definitions} remain true: massive models radiate less efficiently with increasing $\hat\lambda$ at fixed $C_\mathrm{max}$, solitonic models become more efficient with decreasing $\sigma_0$, and all collisions are less efficient than their BH-BH analog.
What is perhaps surprising is how rapidly, at least for solitonic models, the radiative efficiency approaches the BH-BH limit even for compactness values well below the ultracompact 
regime that begins at $C_\mathrm{max} \approx 1/3 $.
Indeed, even for $C_\mathrm{max} = 0.2$, the radiated energy agrees with the BH-BH collision to within 1\% --- comparable to our numerical uncertainty discussed in Appendix~\ref{app: convergence and error analysis} --- for all solitonic models considered.

This suggests that compact BS-BH mergers may produce waveforms significantly degenerate with those from BH-BH binaries, even for BS models that are still well below the ultracompact regime; cf.~also Ref.~\cite{Evstafyeva:2024qvp}.

\subsection{Boost Velocity and Mass Ratio}
\label{sec:boosts}
So far, all the collisions we have presented have been equal-mass, i.e.~$q\defeq M_{\rm BS}/M_{\rm BH}=1$, 
mergers with a fixed initial boost velocity $v = 0.1$.
In this section, we vary these two parameters, determining the effect which each has on our aforementioned observations regarding radiative efficiency.
To this end, we will focus on three of the BS families considered in the previous section: mini BSs, solitonic BSs with $\sigma_0 = 0.15$ (highly radiatively efficient), and massive BSs with $\hat\lambda = 1000$ (radiatively inefficient).

\begin{figure}
    \includegraphics[width=\linewidth]{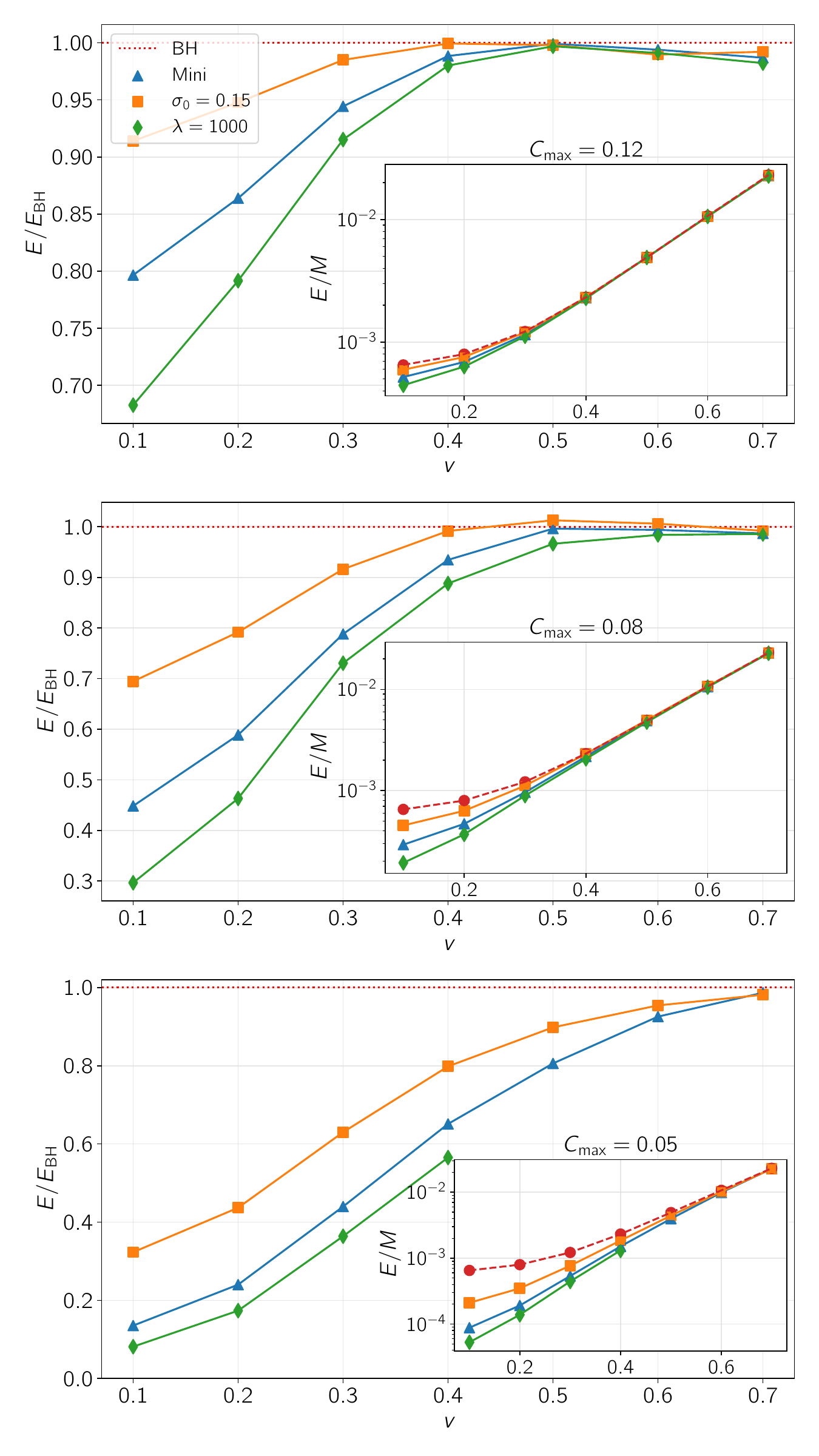}
    \caption{Total energy radiated in GWs for a selection of equal-mass BS-BH head-on collisions using mini, massive, and solitonic models with $C_\mathrm{max} \in \{0.05, 0.08, 0.12\}$, relative to that of an equal-mass BH-BH collision.
    The insets show the raw total energy values $E/M$ for all collisions including BH-BH, for reference. 
    }
    \label{fig:boost_velocities}
\end{figure}

First, we consider the impact of the boost velocity. 
As $v \rightarrow 1$ we expect that ``matter ceases to matter'': the potential (and indeed any property of the compact objects beyond their overall mass) will cease to affect physical observables such as the GW signals \cite{Choptuik:2009ww,Sperhake:2012me,Rezzolla:2012nr}. 
 This expectation, however, does not inform us of the rate at which the relative variation in radiative efficiency observed in the previous section varies with $v.$ A better understanding of this rate will help us to estimate the effect played by this or other matter models in more astrophysically realistic inspiral scenarios.
We systematically explore the matter effects by performing a series of equal-mass BS-BH collisions using models from the three families under consideration with fixed values of the maximum compactness $C_\mathrm{max} \in \{0.05, 0.08, 0.12\}$, and varying the boost velocity for each model from $v = 0.1$ to $v = 0.7$. 

In Fig.~\ref{fig:boost_velocities}, we show the results of these runs in the form of their GW energy 
relative to an equal-mass BH-BH collision.
We see a consistent pattern summarized as follows.
\begin{list}{\rm{(\roman{count})}}{\usecounter{count}
             \labelwidth0.5cm \leftmargin0.7cm \labelsep0.2cm \rightmargin0cm
             \parsep0.5ex plus0.2ex minus0.1ex \itemsep0ex plus0.2ex}
\item The systematic variation of the radiated GW energies with the scalar potential, as seen previously, remains intact across a range of boost velocities.
Specifically, at fixed compactness massive BSs radiate less efficiently than mini BSs, which, in turn, are less efficient than solitonic BSs.
\item
Equal-mass BS-BH collisions remain less efficient than their BH-BH counterparts.
\item
As $v$ increases, the radiative efficiencies of BS-BH collisions converges to the BH-BH value.
\end{list}
The precise rate of this latter convergence varies with the compactness value under consideration.
Generally, however, one might consider its rate surprising: even with $v = 0.7$ we are not yet in the ultrarelativistic regime, so we do not expect the ``matter doesn't matter" conjecture to 
fully hold. 
Nonetheless, at this velocity the discrepancies in radiative efficiency between the objects under consideration barely exceed the $\%$ level.
For more compact models the convergence is still more rapid.

\begin{figure}
    \includegraphics[width=\linewidth]{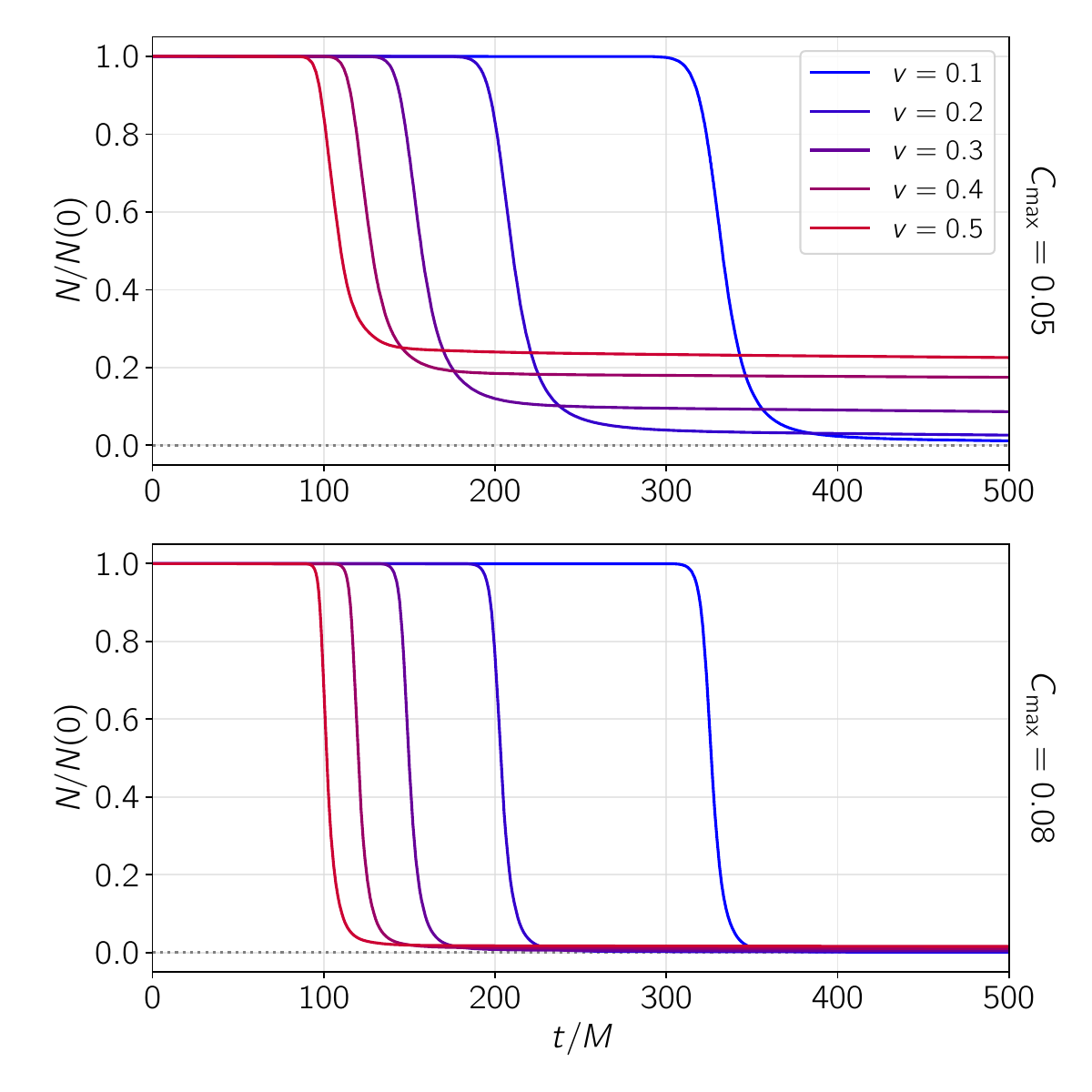}
    \caption{The fraction of the Noether charge $N$ outside the BH horizon is plotted as a function of time for a series of equal-mass mini BS-BH collisions with varying boost velocity for $C_\mathrm{max} = 0.05$ (top) and $C_\mathrm{max} = 0.08$ (bottom).  As the boost velocity is increased, a significantly greater fraction of the scalar matter escapes accretion in the ``fluffy'' $C_{\rm max}=0.05$ case as compared to the more compact $C_{\rm max}=0.08$.
    }
    \label{fig:headon_noether}
\end{figure}

\begin{figure*}
    \includegraphics[width=\linewidth]{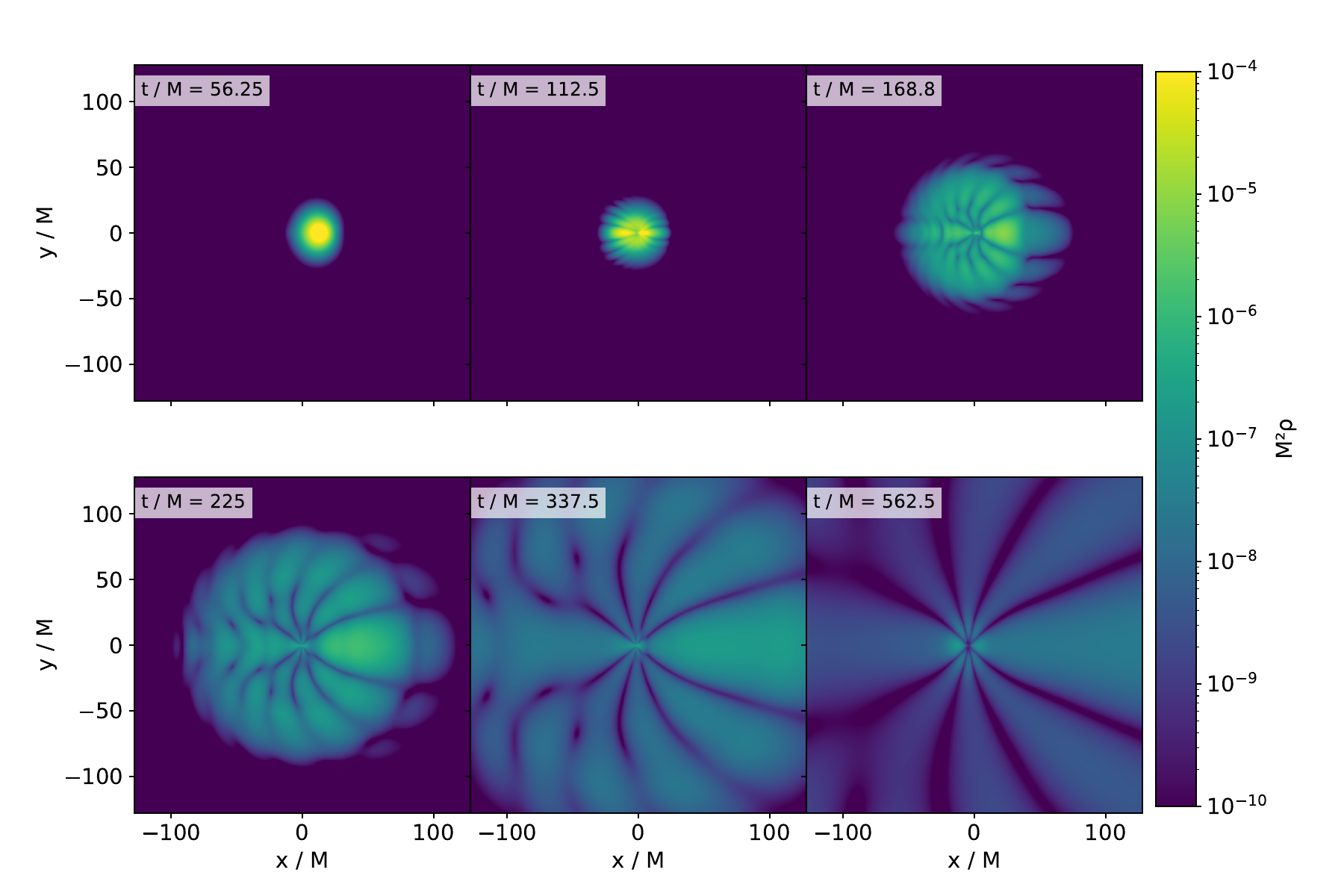}
    \caption{ Snapshots of the energy density $\rho$ --- cf.~Eq.~\eqref{eq:rho_def} ---  during and after the merger for an equal-mass mini BS-BH collision with BS compactness $C_\mathrm{max} = 0.05$ and boost velocity $v = 0.5$.
    }
    \label{fig:rho_v05}
\end{figure*}

Before moving on, we note an additional factor that is likely suppressing GW emission in the $C_\mathrm{max} = 0.05$ case: a significant portion of the scalar matter escapes accretion by the BH. 
To show this, we plot in Fig.~\ref{fig:headon_noether} the Noether charge (i.e.~particle number) over time for two series of mini BS-BH collisions, one with $C_\mathrm{max} = 0.05$, and one with $C_\mathrm{max} = 0.08$ as a reference for the behaviour at higher compactness. 
In the former ``fluffy'' case up to $\sim25\%$ of the scalar matter survives outside the BH well after merger.
This is a significantly larger fraction than has been seen in previous studies of BS-BH collisions~\cite{Cardoso_2022, Zhong_2023} even for mass ratios favoring the BS, $q= M_{\rm BS}/M_{\rm BH}>1$. 
A closer inspection of our data reveals that the remaining scalar matter is mostly dispersed, leaving no clumping remnant, but rather forms a field profile suggestive of an $m = 0$ ``gravitational atom" configuration at late times.  We illustrate this behaviour with snapshots of the energy density profile for the $C_{\rm max}=0.05$, $v=0.5$ binary in Fig.~\ref{fig:rho_v05}. 
We observe qualitatively similar behaviour for all $C_{\rm max}=0.05$ binaries.
The angular index $\ell$ characterizing the late-time configuration, here $\ell = 4$ as seen in the bottom right panel of Fig.~\ref{fig:rho_v05},
varies with the BS compactness but not the boost velocity, and is higher for more compact BSs.

We now turn our attention to the mass ratio $q$. For the remainder of this section, we will return to an initial boost velocity $v = 0.1$. 

\begin{figure*}
    \includegraphics[width=\linewidth]{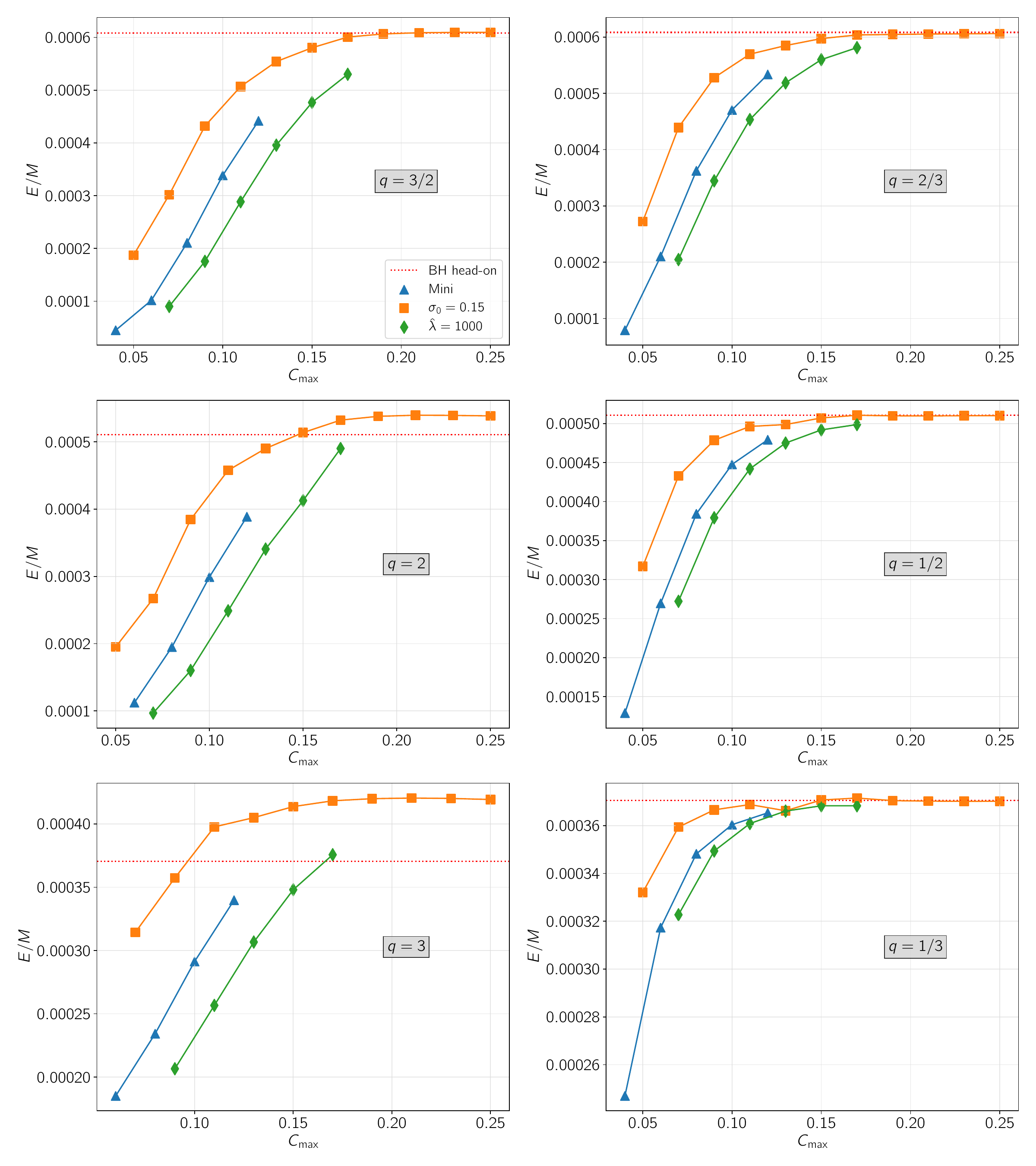}
    \caption{ Total energy emitted in GWs for equal-mass head-on BS-BH collisions against maximum compactness $C_\mathrm{max}$, for a range of mass ratios $q \defeq M_\mathrm{BS} / M_\mathrm{BH}.$ In each case the dotted red line corresponds to a BH-BH collision with the same mass ratio.
    }
    \label{fig:mass_ratios}
\end{figure*}
In Fig.~\ref{fig:mass_ratios} we show the radiated GW energy  as a function of $C_\mathrm{max}$ for our three representative BS families and a variety of mass ratios. 
As before, massive models consistently radiate less  efficiently than mini BSs, which are less efficient than solitonic stars.
Furthermore, when $q < 1$, i.e. when the BH is more massive than the BS, the radiative efficiency continues to approach the BH-BH value as the BS compactness increases.
When the BS mass exceeds that of the BH, however,  a new phenomenon emerges: sufficiently compact BS-BH collisions can emit \textit{more} energy in GWs than the BH-BH collision of equivalent mass ratio.
In this context, we recall that for BH-BH collisions, the radiated energy scales as~\cite{Berti_2007}
\begin{equation}
    \frac{E_\mathrm{BH}}{M} \sim \frac{q^2}{(1 + q)^4}.
\end{equation}
Our results suggest that BS-BH binaries, even with very compact BSs, obey a different scaling law when the BS is more massive than the BH.
A comprehensive exploration of this behaviour will require a large number of additional simulations and is left for future work.

%====================================================================
\subsection{BH-UCO Collisions}
\label{sec:bh_uco}
\begin{table}[t]
    \centering
    \caption{
    Solitonic parameter $\sigma_0$, central amplitude $A_0$, mass $M_\mathrm{BS}$, Noether charge $N$, radius $r_{99}$, and maximum compactness $C_\mathrm{max}$ for the two ultracompact BSs we consider in this work.
    }
    \begin{ruledtabular}
    \begin{tabular}{lcccccc}
        Model & $\sigma_0$ & $A_0$ & $\mu M_\mathrm{BS}$ & $\mu^2 N$ & $\mu r_{99}$ & $C_\mathrm{max}$ \\ \hline
        \texttt{S06A044} & 0.06 & 0.044 & 10.43 & 65.9 & 31.7 & 0.33 \\ 
        \texttt{S08A06} & 0.08 & 0.06 & 5.65 & 26.1 & 17.9 & 0.31 
    \end{tabular}
     \end{ruledtabular}
    \label{tab:thinshell_info}
\end{table}

Finally, we present a selection of BS-BH collisions in which the BS is ultracompact, i.e.~possesses a pair of light rings, one unstable and one stable. 
To this end we use the \textit{thin-shell} models, obtained for a solitonic potential~\eqref{eq:solitonic} with sufficiently small $\sigma_0$; cf.~Ref.~\cite{Marks_2025} where such BS models have
been demonstrated to be dynamically stable on long timescales.
We summarise the properties of our specific thin-shell models in Table~\ref{tab:thinshell_info}.
We also note that simulating collisions involving these models is computationally much more challenging  compared to the BS binaries we have considered previously.
This is because of the steep gradient in the scalar field profile at relatively large radius, forming the thin shell from which these models take their name. To correctly capture their dynamics, this shell region must be well-resolved. For equal-mass collisions of our specific models, we therefore need to increase the
resolution on the finest grid level from $\du x = M/64$ to $\du x = M / 80$ or higher to maintain a percent level accuracy. 
The computational cost is somewhat mitigated by {\sc grchombo}'s capacity for AMR, which can concentrate refinement levels in regions of steep gradients, thus avoiding unnecessarily high resolutions in the comparatively smooth BS interior.

\begin{figure}
    \includegraphics[width=\linewidth]{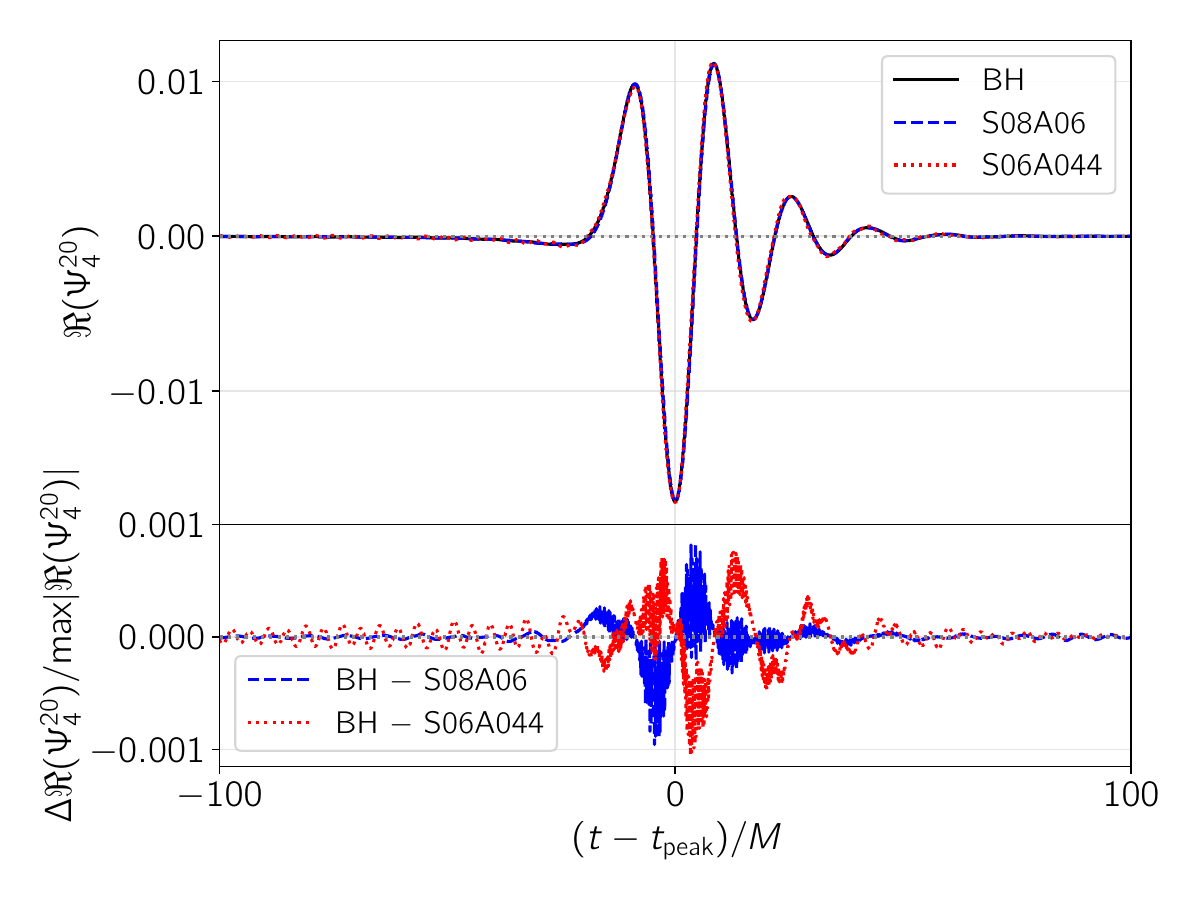}
    \caption{ Real part of the dominant $(2,0)$ mode of the Newman-Penrose scalar $\Psi_4$ over time for equal-mass head-on collisions of a Schwarzschild BH with another BH or either of the thin-shell BSs presented in Table~\ref{tab:thinshell_info}. In the lower panel we show differences between the BH-BH and BH-UCO collisions, normalized by the maximum amplitude in the BH-BH case. The oscillatory discrepancies at early and late times in the \texttt{S06A044} case are likely due to larger numerical errors in the initial data for this more extreme model. 
    }
    \label{fig:thinshell_waveforms}
\end{figure}

We first perform head-on collisions of either of the two models in Table~\ref{tab:thinshell_info} with an equal-mass BH.
In Fig.~\ref{fig:thinshell_waveforms}, we compare the dominant $(2, 0)$ modes of the Newman-Penrose scalar $\Psi_4$ for these runs to those of an analogous BH-BH collision, showing that these are effectively indistinguishable within our numerical uncertainty. 
Given our previous results concerning non-ultracompact BSs, this is hardly surprising, but it strengthens the numerical evidence for another respect in which UCOs mimick BHs: their GW emission in binaries.

\begin{figure*}
    \includegraphics[width=\linewidth]{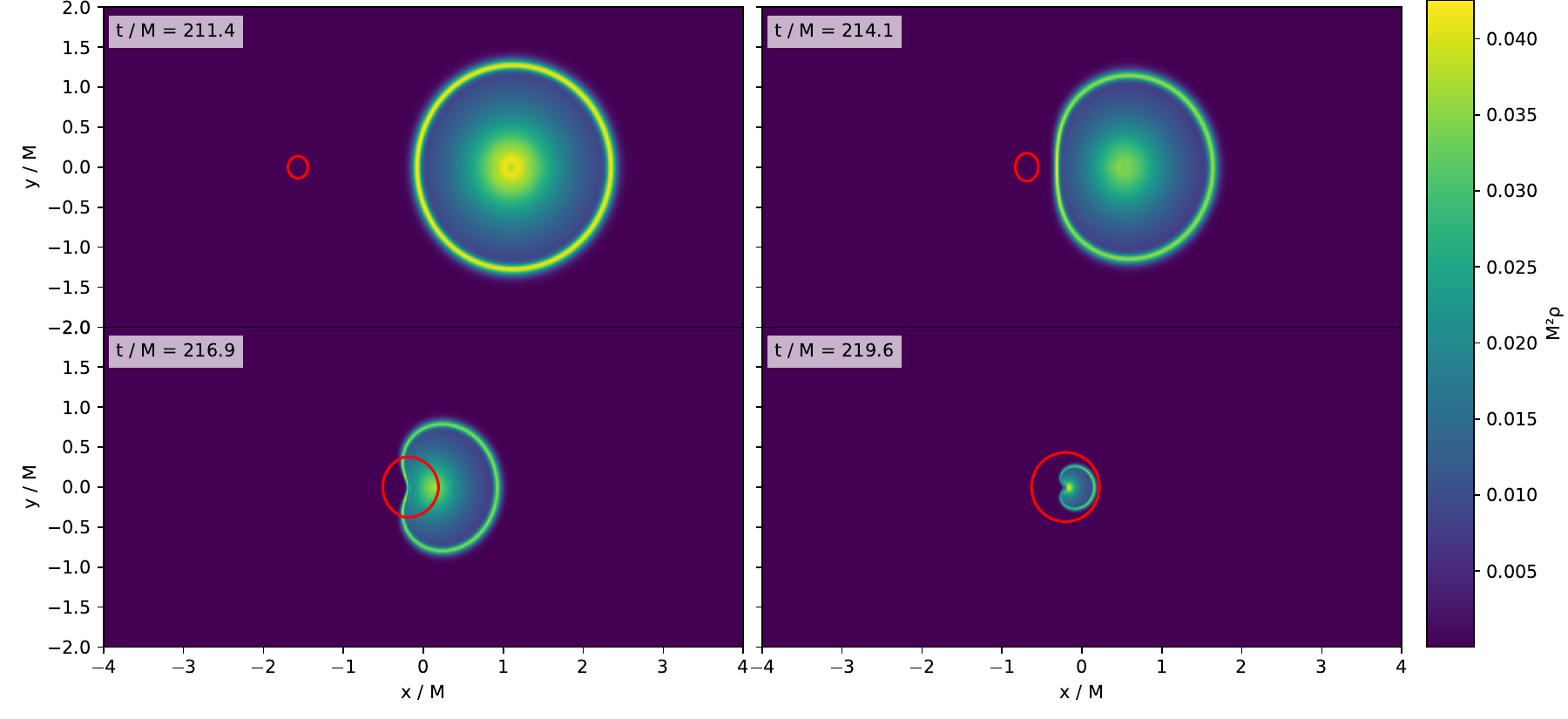}
    \caption{ Snapshots of the energy density $\rho$ before, during and after merger for BS \texttt{S08A06} colliding with a less massive BH with mass ratio $q = 3$. The contour in red corresponds to conformal factor $\chi = \left[\det\gamma\right]^{-1/3}= 1/16,$ the value at the event horizon for a Schwarzschild BH in isotropic coordinates, and therefore gives a rough estimate for the horizon.
    }
    \label{fig:rho_ts}
\end{figure*}

We also perform runs with mass ratio $q = 3$ (with the BS more massive). 
In Fig.~\ref{fig:rho_ts} we show a series of representative snapshots of the energy density profile around merger.
Note that the ``shell-like'' structure of the energy density profile, characteristic of these ultracompact BSs, remains intact even under the large tidal disruption associated with merger.
This seeming robustness of the shell profile is commensurate with the results of Ref.~\cite{Marks_2025}, where we conjectured that it may help this class of UCOs evade the potential instability associated with the presence of a stable light ring.

%=============================================================================
\section{Inspirals}
\label{sec:inspirals}
We now turn our attention to a first study of BS-BH binary inspirals, aiming to determine to what extent the results we found in Sec.~\ref{sec:headon} generalize beyond the case of head-on collisions.
To this end, we consider three equal-mass inspirals, with identical configuration apart from the use of a different BS model: one mini, one massive with $\hat\lambda = 100$, and one solitonic with $\sigma_0 = 0.15$. The configuration we use corresponds to the quasi-circular binary parameter choices \texttt{q1-d12} in Ref.~\cite{Radia_2021}, where it was used to benchmark {\sc grchombo}'s capacity to accurately evolve BH-BH binaries. 
As before, our chosen models share a common value of the maximum compactness $C_\mathrm{max} = 0.1$.
Other global quantities characterizing the BS constituents are summarized in Table~\ref{tab:inspiral_info}.
For all these configurations, we use {\sc twopunctures}-corrected initial data as described in section~\ref{sec:initial_data}.

We can predict some qualitative features of our BH-BS inspirals based on an analogy with previous work done on nonspinning BH-NS inspirals~\cite{Duez:2009yy, Kyutoku_2010}; cf.~Ref~\cite{Kyutoku:2021icp} for a review. 
Such NS studies suggest that BS-BH inspirals  fall into one of two  categories: (i) the BS is tidally disrupted before merger, introducing a cutoff frequency into the resulting GW signal, and (ii) plunges where no appreciable tidal disruption occurs, but the BS is instead swallowed by the BH in one go.
In the NS case, tidal disruption occurs when the radius $r_{\rm ms}$ at which mass-shedding begins, is larger than the radius of the BH's innermost stable circular orbit, $r_\mathrm{ISCO}$. 
A Newtonian approximation estimates this ratio (for a nonspinning BH) as~\cite{Kyutoku:2021icp}
\begin{equation}
    \!\!\!\!\frac{r_\mathrm{ms}}{r_\mathrm{ISCO}} = \frac{2^{\frac13} c_\mathrm{R} }{6 C_\mathrm{NS}} \left( \frac{M_\mathrm{NS}}{M_\mathrm{BH}}\right)^{\frac23}
    \!\!\approx
    0.21 \frac{c_\mathrm{R}}{C_\mathrm{NS}} \left( \frac{M_\mathrm{NS}}{M_\mathrm{BH}}\right)^{\frac23}
    , \label{eq:tidal_disruption}
\end{equation}
where $C_\mathrm{NS}$ is the NS compactness and $c_\mathrm{R} \ge 1$ represents the degree of tidal elongation of the NS, the precise extent of which is model-dependent. Following a tidal disruption event, at least half the NS material is rapidly accreted by the BH, while the remainder forms an accretion disk; the underlying mechanisms are
comparable to those originally proposed for ordinary stars in Ref.~\cite{Rees1988TidalDO}.

In view of these findings, a naive application of Eq.~\eqref{eq:tidal_disruption} to our equal-mass BS-BH configuration with $C_\mathrm{max} = 0.1$ --- which is a generous definition of the compactness --- would imply tidal disruption irrespective of the scalar potential.  
Of course, Eq.~(\ref{eq:tidal_disruption}) is only an approximation: (i) in the NS case
the surface self-gravity ought to be treated general-relativistically, and (ii) in the BS case, effects of the scalar interaction are not accounted for at all.
As in the BH-NS case, a comprehensive understanding of the precise nature of tidal disruption for BS-BH binaries must be studied numerically.

\begin{table}[t]
    \centering
    \caption{Scalar potential, central amplitude $A_0$, Noether charge $N$, binding energy $E_B$, and compactness $C_0 \defeq M_\mathrm{BS}/r_{99}$ for the three BSs used in equal mass BS-BH inspirals in this section. All three models have $C_\mathrm{max} = 0.1$.
    }
    \begin{ruledtabular}
    \begin{tabular}{lcccc}
        Potential & $A_0$ & $\mu^2 N$ & $\mu E_B$ & $C_0$\\ \hline
        Mini & 0.0598 & 0.644 & -0.0188 & 0.0683 \\ 
        Massive, $\hat\lambda = 100$ & 0.0347 & 0.943 & -0.0381 & 0.0745   \\  
        Solitonic, $\sigma_0 = 0.15$ & 0.116 & 0.542 & -0.124 & 0.0805  
    \end{tabular}
    \end{ruledtabular}
    \label{tab:inspiral_info}
\end{table}
\begin{figure*}
    \includegraphics[width=\linewidth]{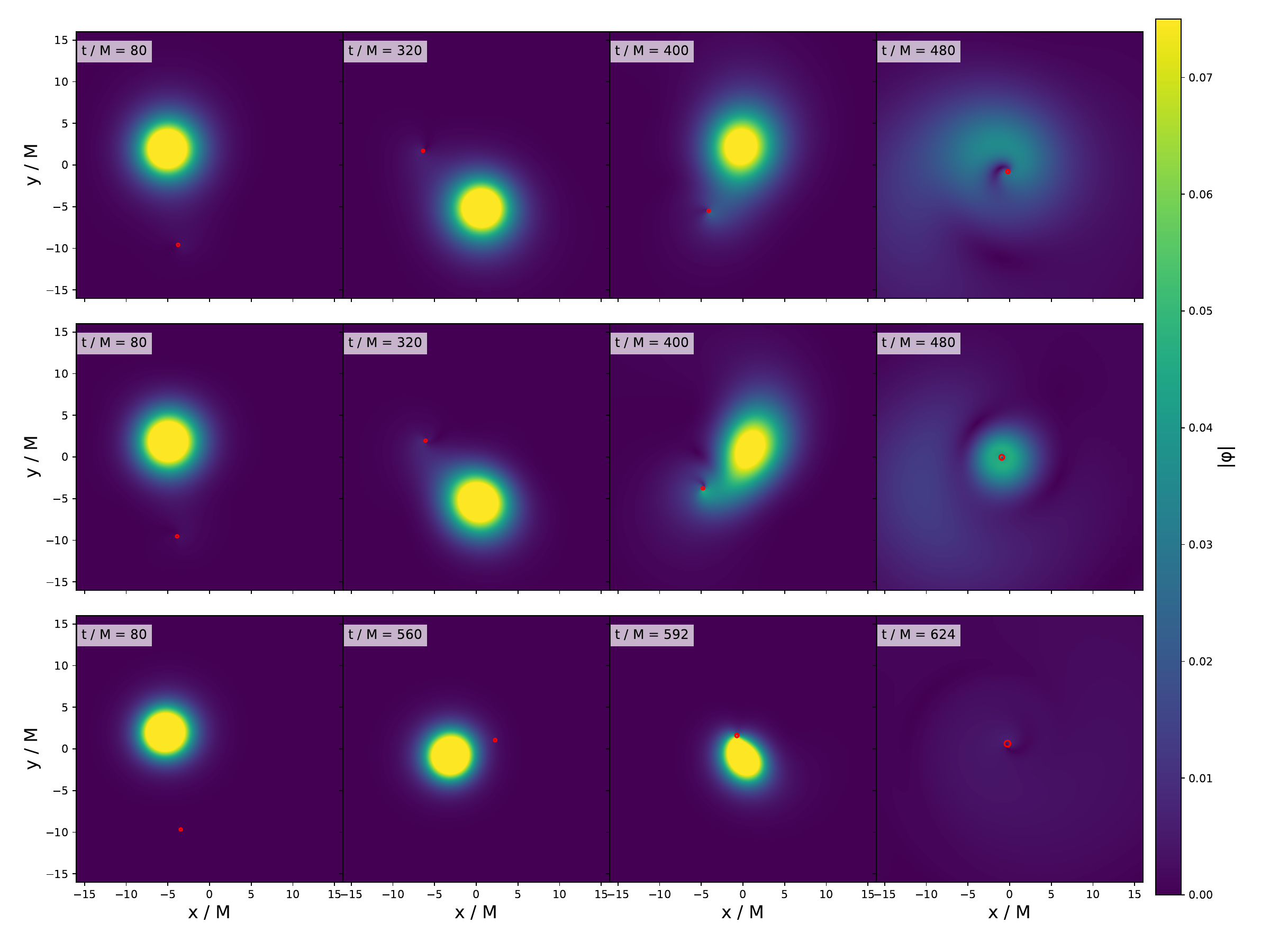}
    \caption{Snapshots of the scalar-field amplitude (normalized by the central amplitude of each BS in equilibrium) before, during and after merger for our mini (top), massive (middle) and solitonic (bottom) inspiral configurations (cf. Table~\ref{tab:inspiral_info}). The contour in red corresponds to conformal factor $\chi = 1/16,$ the value at the event horizon for a Schwarzschild BH in isotropic coordinates, and therefore gives a rough estimate for the horizon. In the solitonic case, closer inspection reveals a small amount of ejecta matter, but only a negligible amount of scalar matter surrounding the BH after merger.
    }
    \label{fig:inspiral_profiles}
\end{figure*}

\begin{figure}
    \includegraphics[width=\linewidth]{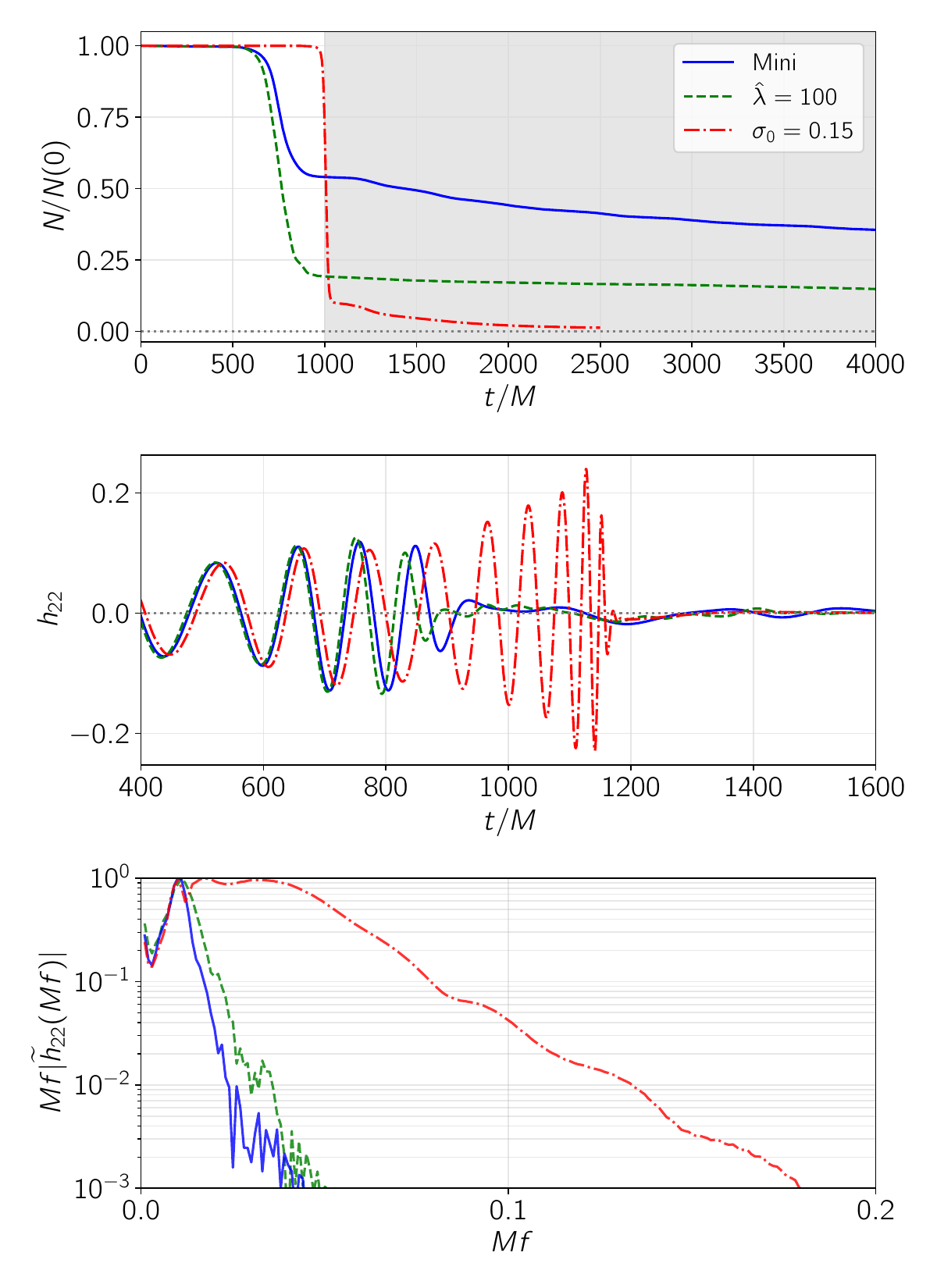}
    \caption{ \textit{Top:} Fraction of the non-accreted Noether charge over time. The shaded region corresponds to the slow accretion phase, after the initial prompt one. \textit{Center:} Dominant (2,2) mode of the GW strain $h$ over time. \textit{Bottom:} Power spectra of the waveform above, normalized to unity, showing that the mini and massive signals are ``cut off" at significantly lower frequencies than their solitonic counterpart. 
    }
    \label{fig:insp_quantities}
\end{figure}

To this end, we now present the results of our three inspiral configurations. 
In Fig.~\ref{fig:inspiral_profiles}, we show snapshots of the scalar amplitude in the orbital plane for each binary.
Strikingly, while the mini and massive models display the expected tidal disruption before merger, the solitonic model does not.
We also see that an appreciable quantity of scalar matter remains outside the BH in both tidally-disrupted scenarios.
In Fig.~\ref{fig:insp_quantities}, we show the non-accreted Noether charge and GW signals for each configuration. 
Interestingly, in the mini BS case the amount of Noether charge initially accreted after tidal disruption is very close to the 50\% predicted by a naive Newtonian calculation~\cite{Rees1988TidalDO}, but for the massive BS it is closer to 80\%.
This occurs despite the fact that both BSs are tidally disrupted at roughly the same time.
Furthermore, after the initial prompt accretion phase, the scalar matter in the mini case is actually accreted more rapidly: the Noether charge is reduced by $\sim$34\% between $t = 1000M$ and $t = 4000M$ --- corresponding to the shaded region in Fig.~\ref{fig:insp_quantities} --- for the mini potential, but only by $\sim$23\% for the massive potential.
This is consistent with recent semi-analytical results suggesting that the accretion rate of a massive scalar field with quartic self-interaction decreases with the interaction strength \cite{Glavan_2026, Feng:2021qkj}.
We also see that, as expected, the BS tidal disruptions are associated with a cutoff frequency in (the inspiral part of) the GW signal, relative to the solitonic case which does not undergo tidal disruption before merger.

Our results suggest that the phenomenology of BS tidal disruption can be partially understood in terms of our rough potential-EoS analogy.
In this context, the binding energy is likely a relevant factor:
we can attribute the resilience against disruption of our solitonic model to its large (negative) binding energy, corresponding to a strongly gravitationally bound system
(cf.~Table~\ref{tab:inspiral_info}).
This new argument, however, does not explain why the massive model is even more vulnerable to tidal deformation than the mini-BS model, despite also having stronger
gravitational binding.
Nor is this explained by an alternative compactness definition such as $C_0,$ because the value of $C_0$ for the massive model falls between that of the mini and solitonic models.

{Of course, these early results do not guarantee that the same resistance to tidal disruption shown by the solitonic model will persist across a broader range of mass ratios and orbital separations. Given that the equal-mass configuration studied here is so far into the regime where tidal disruption is expected, however, we anticipate that the general pattern of greater resilience will remain.}

%=============================================================================
\section{Conclusions}
\label{sec:conclusion}
We have presented an extended study of the dynamics of BS-BH binaries, focusing on the non-spinning and comparable-mass regime and on the impact of varying models for the scalar-field potential.
To this end we have demonstrated the importance of using equilibrated initial data in dynamical evolutions, and suggested measures for further reducing the initial data's constraint violations.

By simulating BH-BS head-on collisions across a wide range of mass ratios $q$, boost velocities $v$ and scalar potentials, we have shown that the matter model used for the BS remains significant for their merger dynamics with BHs: for any plausible definition of the BS compactness, knowledge of this compactness and the mass ratio alone is not sufficient to predict the resulting GW emission with good accuracy.
This enriches the GW phenomenology expected from BS-BH binary systems, and, thus, may limit the scope of model-agnostic GW searches for ECOs.
Our results also suggest a loose correspondence between common choices for the scalar potential of BSs and NS equations of state: broadly speaking, a massive (solitonic) potential corresponds to a stiff (soft) EOS. This analogy excludes, however, the extreme
regime of thin-shell BSs obtained for solitonic potentials
with small $\sigma_0$; for these the 6th-order term in the 
potential becomes dominant, drastically changing its character
from ``soft'' to ``very stiff.''
Furthermore, we have presented the first collisions of a BH and such a thin-shell UCO, showing that the resulting GW signal effectively mimics that of the analogous BH-BH collision, and that the characteristic \textit{thin-shell} structure of our UCOs' energy-density profile retains its integrity almost all the way up to merger.

Our preliminary investigation of BS-BH inspirals has further underscored the impact of the potential function, and suggests both parallels with and possible departures from the behavior of analogous BH-NS mergers.
In particular, we have seen that an appropriate choice of potential can cause the BS to resist tidal deformation at a compactness and mass ratio  where NS analogs suggest disruption.
It will be interesting to determine whether this can be explained in terms of BS tidal deformability calculated \textit{a priori} as for example in Ref.~\cite{Sennett_2017}, or via an effective-one-body model incorporating scalar matter \cite{Damour_2025}.

In general, a more systematic study of the prospects for distinguishing BSs and other ECOs using GW astronomy is needed. 
For instance, while we have compared BS-BH waveforms to their BH-BH counterparts in this work, we have not considered any possible degeneracies with BH-NS waveforms, which have recently received fresh numerical attention in the near-equal mass ratio regime \cite{Markin_2026}.
An understanding of the extent of these degeneracies and possible means to break them will likely aid GW searches for ECOs. 
Due to the high computational cost of dynamical evolutions using full NR, it is likely that  surrogate models, which have recently been constructed for head-on collisions of Proca stars \cite{Luna_2024}, will lead the way for ECO template banks, with a smaller set of NR simulations helping to inform on what features they must take into account.

%=============================================================================

\begin{acknowledgments} G.A.M. is supported by the Cambridge Trust at the University of Cambridge, and would like to thank Tamara Evstafyeva and Christopher J. Moore for helpful conversations. S.J.S. is supported by the Centre for Doctoral
Training at the University of Cambridge funded through STFC (ST/W006812/1, 2882538). This
work has been supported by STFC Research Grant No. ST/V005669/1.
We acknowledge support by the NSF Grant Nos.~PHY-090003,~PHY-2513337
and AST-2307146, DiRAC projects ACTP284 and ACTP238, STFC capital
Grants Nos.~ST/P002307/1,~ST/R002452/1,~ST/I006285/1 and ST/V005618/1,
STFC operations Grant No.~ST/R00689X/1. Computations were done on
the CSD3, Swirles and Fawcett (Cambridge), Cosma (Durham), Stampede3 (TACC) and Expanse (SDSC) clusters.
\end{acknowledgments}

\bibliography{ref}

\appendix
\newcommand{\cgamma}{\tilde{\gamma}}
\newcommand{\cGamma}{\tilde{\Gamma}}
\newcommand{\hGamma}{\hat{\Gamma}}

\begin{figure}
    \includegraphics[width=\linewidth]{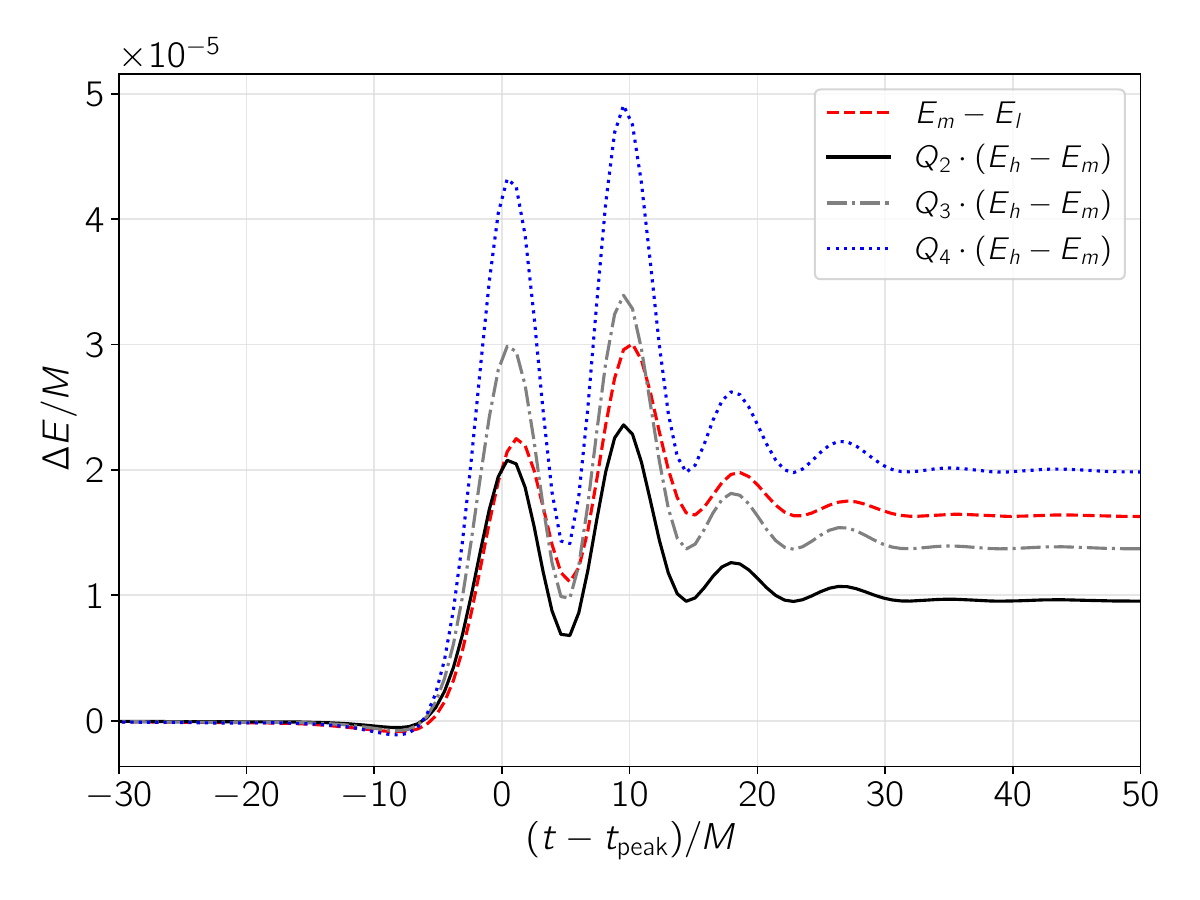}
    \caption{ Resolution study of the total energy emitted as GWs over time for an equal-mass BS-BH head-on collision using a solitonic BS with $\sigma_0 = 0.2$ and central amplitude $A = 0.157$, using resolutions $\Delta x_l = M/32, \; \Delta x_m = M/48, \; \Delta x_h = M/64$. The black, gray and blue curves show the differences between high- and medium-resolution runs amplified by the $p$th-order convergence factor $Q_p \defeq (\Delta x_m^p - \Delta x_l^p) / (\Delta x_h^p - \Delta x_m^p)$ for $p = 2,~3$ and $p = 4$, respectively. Comparison with the difference between
    low and medium resolution (red dashed curve) yields convergence of approximately second-order at early times and between third and fourth order at late times.
    }
    \label{fig:E_convergence}
\end{figure}

\section{Evolution System} \label{sec:evolution_equations}
In this appendix, we present the evolution equations used in this work, based on the CCZ4 formulation of numerical relativity \cite{Alic:2013xsa}.
This formulation is obtained by appending to the Einstein-Hilbert action a term involving a new vector field $Z^\mu,$
\begin{equation}
    S_Z = \frac{1}{16\pi}\int\du^{4}x\,\sqrt{-g} \nabla_\mu Z^\mu\,,
\end{equation}
such that physical solutions satisfy $Z^\mu = 0$. The benefit of this addition is that it acts like a damping term for the constraint equations, reducing drifts away from the physical, constraint-satisfying spacetime.

In the following we denote by $\alpha,\beta^i, n^\mu$ the lapse, shift vector, and timelike unit normal vector, respectively.
The evolution variables for the geometry are then given by the conformal factor $\chi \defeq (\det \gamma)^{-1/3},$ the conformally rescaled metric $\tilde{\gamma}_{ij} \defeq \chi\gamma_{ij},$ the trace of the extrinsic curvature $K,$ the conformally rescaled trace-free extrinsic curvature $\tilde{A}_{ij} \defeq \chi\left(K_{ij} -\frac{1}{3}K\gamma_{ij} \right)$, the Z4 variables $\Theta \defeq -n_\mu Z^\mu$ and $\Theta^i \defeq Z^i - \Theta n^i$, and the contracted conformal Christoffel symbols $\hat{\Gamma}^i \defeq \tilde{\gamma}^{jk} \tilde{\Gamma}^{i}_{jk} + \frac{2}{\chi}\Theta^i .$
We model the scalar field in terms of two first-order-in-time differential equations obtained by introducing the momentum $\Pi \defeq -\frac{1}{2\alpha}(\partial_t \varphi - \beta^k \partial_k \varphi).$
The energy density, momentum density, and stress tensor are then given by
\begin{align}
    \rho = \, & 2\Pi\bar\Pi  + \frac{1}{2}\partial_m\varphi\partial^m\bar\varphi + \frac{1}{2}V, \label{eq:rho_def} \\
    j_i = \, &\bar\Pi\partial_i\varphi + \Pi\partial_i\bar\varphi, \\
    S_{ij} = \, & \partial_{(i}\varphi\partial_{j)}\bar\varphi - \frac{1}{2}\gamma_{ij} \left(\partial_m\varphi\partial^m\bar\varphi - 4 \Pi\bar\Pi  +V \right).
\end{align}
We also use $\mathcal{R}_{ij}$ and $\mathcal{R}$ to denote the spatial Ricci tensor and scalar, $D_i$ for the spatial covariant derivatives, and the superscript $\mathrm{TF}$ to denote trace-free parts.
The lapse and shift are evolved according to the standard moving puncture gauge conditions \cite{Campanelli:2005dd, Baker:2005vv}. 
Finally, introducing the damping parameters $\kappa_1$, $\kappa_2,$ $\kappa_3$ and $\eta$, the set of evolution equations used for our BS-BH binaries is given by,

\begin{widetext}
\begin{align}
        \partial_t\chi = \, & \beta^m\partial_m\chi + \frac23\chi\left(\alpha K - \partial_m\beta^m\right), \\
        \partial_t\cgamma_{ij} = \, & \beta^m\partial_m\cgamma_{ij} + 2\cgamma_{m(i}\partial_ {j)} \beta^m - \frac23 \cgamma_{ij} \partial_m\beta^m - 2\alpha\tilde{A}_{ij}, \\
        \partial_tK = \, & \partial_m\beta^m K - D^mD_m\alpha + \alpha \bigg[ \mathcal{R} + K(K - 2\Theta) + 2D_m\Theta^m - 3\kappa_1(1+\kappa_2)\Theta + 4\pi(S - 3\rho)\bigg], \\
        \partial_t\tilde A_{ij} = \, & \beta^m\partial_m \tilde A_{ij} + 2\tilde A_{m(i}\partial_{j)}\beta^m - \frac23 \tilde A_{ij}\partial_m\beta^m - 2\alpha\tilde A_{im}\tilde A^m_{\;\;j}  + \alpha\tilde A_{ij}(K - 2\Theta) 
        \\ \nonumber & - \chi \bigg[\alpha(\mathcal{R}_{ij} - 8\pi S_{ij} + 2D_{(i}\Theta_{j)} ) - D_iD_j\alpha   \bigg]^\mathrm{TF}, \\
        \partial_t\hGamma^i = \, & \beta^m\partial_m\hGamma^i + \cgamma^{mn}\partial_m\partial_n\beta^i + \frac13\cgamma^{ik}\partial_k\partial_m\beta^m + \frac23\hGamma^i\partial_m\beta^m - 2\partial_m\alpha(\tilde A^{im} + \Theta \cgamma^{im}) - \hGamma^m\partial_m\beta^i + 2\alpha \tilde A^{mn}\cGamma^i_{mn} \\
        \nonumber& - 3\alpha\tilde A^{im}\frac{\partial_m\chi}{\chi} + 2\alpha\cgamma^{im}\left(\partial_m\Theta - \frac23 \partial_m K \right) - \frac{2\kappa_1\alpha}{\chi}\Theta^i - \frac{4\alpha K}{3\chi}\Theta^i - \frac{16\pi\alpha}{\chi}j^i + \frac{\kappa_3}{\chi}\left(2\partial_k\beta^k \Theta^i - 3\Theta^k\partial_k\beta^i \right), 
        \\
        \partial_t\Theta = \, & \beta^m\partial_m\Theta + \frac\alpha 2\left[\mathcal{R} + 2D_m\Theta^m - \tilde A_{mn}\tilde A^{mn} + \frac23 K^2 - 2K\Theta- 2\Theta^m \frac{\partial_m\alpha}{\alpha} - \kappa_1(4 + 2\kappa_2)\Theta - 16\pi\rho \right],
                \\
        \partial_t\varphi = \, & \beta^m\partial_m\varphi - 2\alpha\Pi  , \\
        \partial_t\Pi  = \, & \beta^m\partial_m \Pi + \alpha K\Pi + \frac12\alpha\varphi V - \frac12\chi\cgamma^{mn}\left( \partial_m\varphi\partial_n\alpha + \alpha \tilde D_m \tilde D_n\varphi \right) + \frac14\alpha\cgamma^{mn}\partial_m\varphi\partial_n\chi,  \\
        \partial_t\alpha = \, & \beta^m\partial_m \alpha - 2\alpha K, \\
        \partial_t\beta^i = \, & \beta^m\partial_m \beta^i + \frac{3}{4}\cGamma^i - \eta\beta^i.
    \end{align}
\end{widetext}

{
\section{More on the End-states}
In this appendix, we present additional results pertaining to the gravitational atom configurations that serve as end-states of our BS-BH mergers.
First, in Fig.~\ref{fig:phi_v05} we show a representative late-time profile of the real part of $\varphi$ for one of our head-on collisions with a boost velocity of $v = 0.5$.
Notice the presence of a contour enclosing the BH at coordinate radius $r \sim 11M$ for which the real part of the scalar field is uniformly zero.
Recall also that our head-on collisions are performed in enforced (and twist-free) axisymmetry.
Thus, the winding number of $\mathrm{Re}(\varphi)$ around the puncture is zero, justifying our identification of this configuration as an $m = 0$ gravitational atom.

\begin{figure}
    \includegraphics[width=\linewidth]{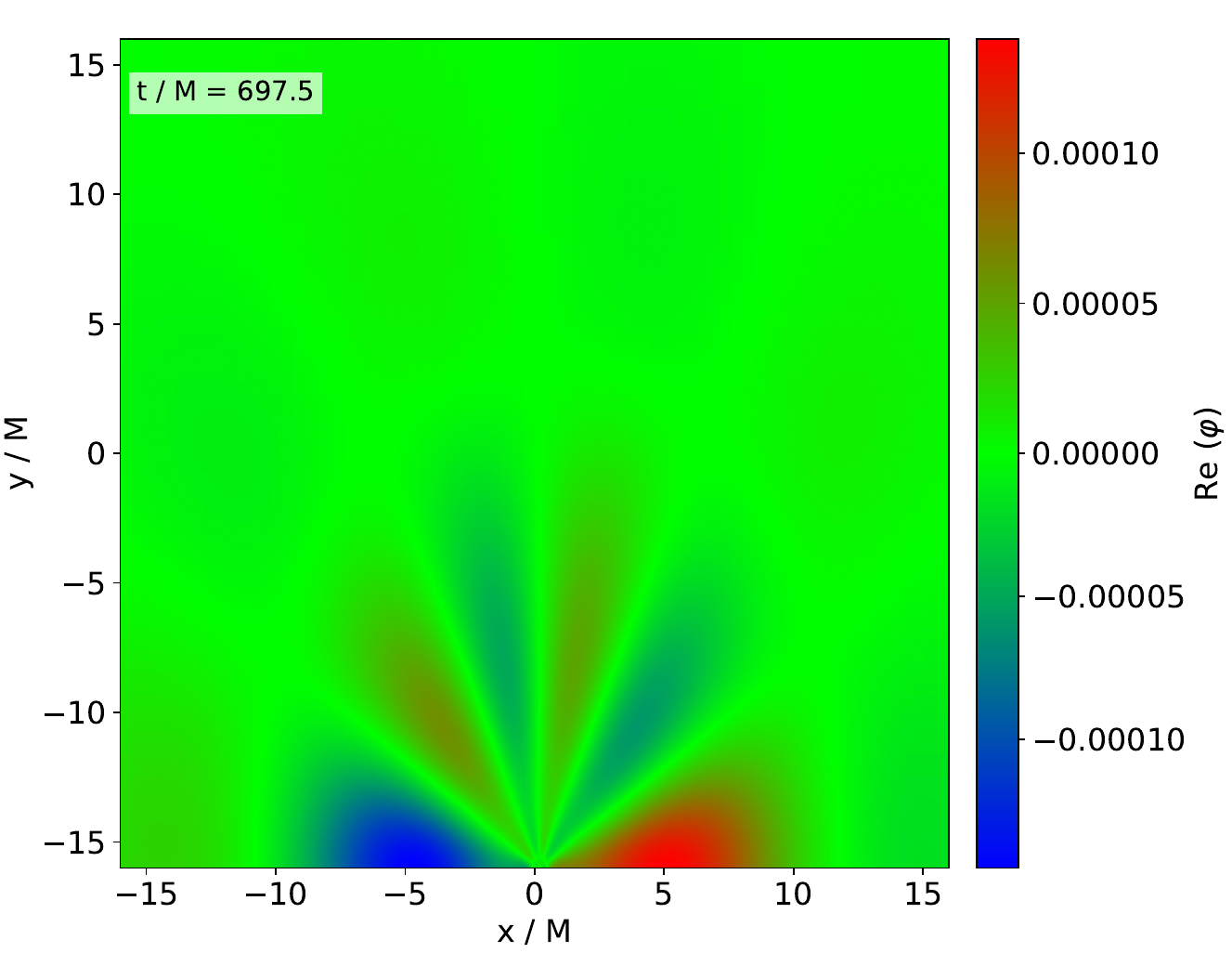}
    \caption{ 
    Late-time profile of the real part of the scalar field $\varphi$ for an equal-mass mini BS-BH collision with BS compactness $C_\mathrm{max} = 0.08$ and boost velocity $v = 0.5$.
    }
    \label{fig:phi_v05}
\end{figure}

We also briefly discuss in more detail the end-states of our BS-BH inspiral configurations, obtained by running an apparent horizon finder on each of our configurations at late times.
In Table~\ref{tab:final_BHs}, we give the final irreducible masses and spins for each of the inspirals we consider, comparing to the analogous result for a quasicircular inspiral of two equal-mass Schwarzschild BHs.
Interestingly, the final spins all lie very close to the BH-BH value, although as expected the horizon masses are significantly smaller for those configurations which undergo tidal disruption and leave a corotating cloud of scalar matter.

\begin{table}[t]
    \centering
    \caption{
    Final masses $M_f$ and dimensionless spins $\chi_f$ of the BH for each of our inspiral configurations, including data for the case of two Schwarzschild BHs (taken from Ref.~\cite{Scheel_2009}) as a reference. See Table~\ref{tab:inspiral_info} for details of the BS models in these binaries.
    }
    \begin{ruledtabular}
    \begin{tabular}{lcc}
        Model & $M_f / M$ & $\chi_f$  \\ \hline
        Schwarzschild BH & 0.95 & 0.69 \\ 
        Mini BS &0.71 & 0.69 \\ 
        Massive BS & 0.78 & 0.68 \\ 
        Solitonic BS & 0.91 & 0.71 
    \end{tabular}
    \end{ruledtabular}
    \label{tab:final_BHs}
\end{table}
}
\section{Convergence and Error Analysis} 
\label{app: convergence and error analysis}
For validation of our numerical results, we analyze in this
appendix the convergence properties of our code and derive the
numerical uncertainties due to discretization and finite
extraction radius.

\begin{figure}
    \includegraphics[width=\linewidth]{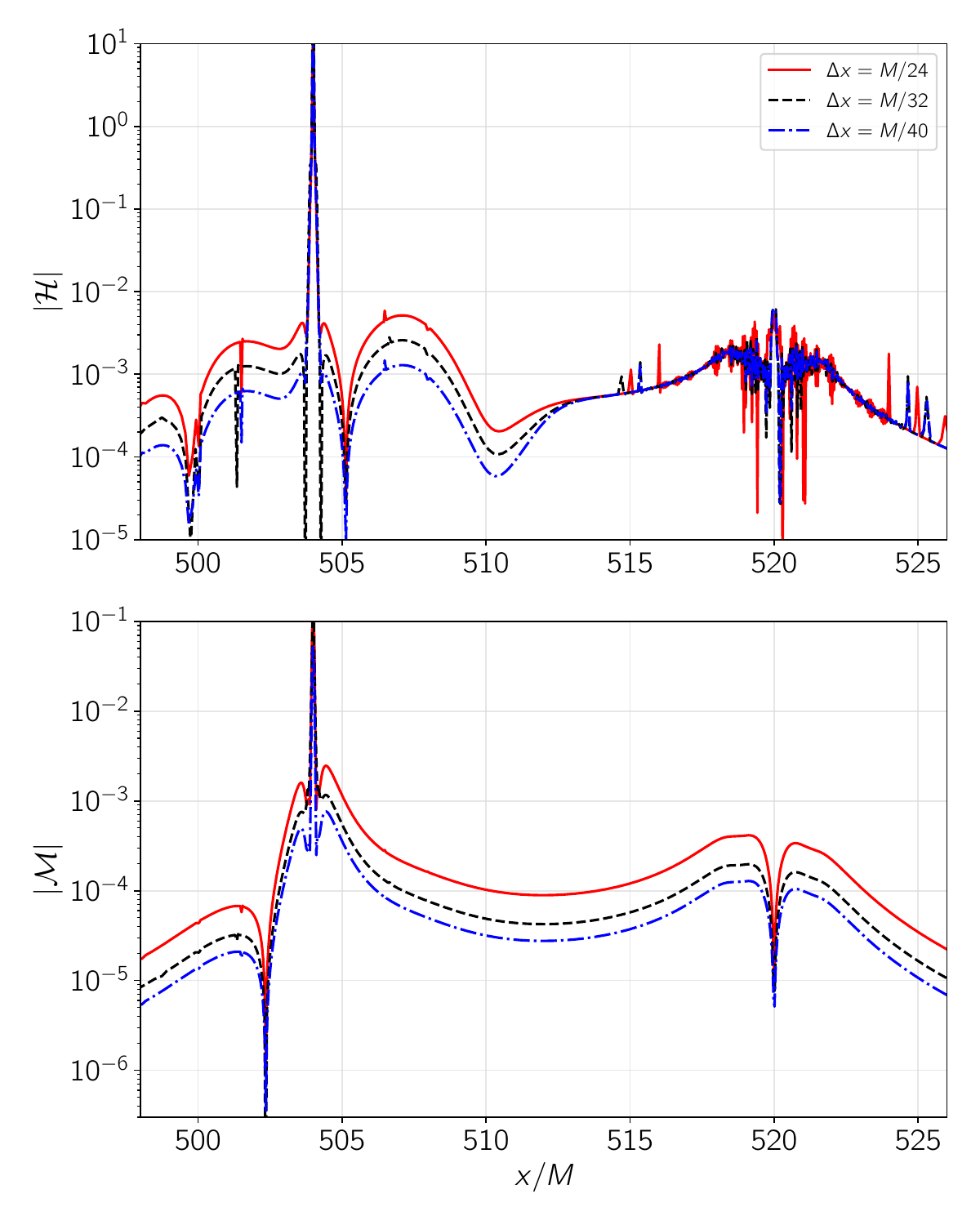}
    \caption{
    The Hamiltonian (top) and
    magnitude $\mathcal{M}= \sqrt{\mathcal{M}_x^2+\mathcal{M}_y^2+\mathcal{M}_z^2}$
    of the momentum (bottom) constraint violations for the {\sc TwoPunctures}-corrected initial data, using the same configuration as in Fig.~\ref{fig:constraint_lineouts} computed at three resolutions.
    }
    \label{fig:convergence_lineouts}
\end{figure}

{
First, we discuss the convergence properties of our {\sc TwoPunctures}-corrected initial data.
A useful representation is provided in Fig.~\ref{fig:convergence_lineouts}, where we show constraint violations along the collision axis for the configuration we used in Fig.~\ref{fig:constraint_lineouts}.
For the Hamiltonian constraint, convergence is consistent with second- to fourth-order around the BH, which is expected since we use constraint-solved data there.
Around the BS, we observe convergence towards a finite residual constraint violation as the metric correction used there is not fully constraint-satisfying, although in practice we find that our CCZ4 evolutions quickly damp this away.
For the momentum constraint, convergence remains consistent with second- to fourth-order throughout the domain, at least at the resolutions we use for inspirals.}

Next, we assess the numerical error due to the finite-differencing approximation. For this purpose, we have
evolved the equal-mass head-on collision of a BH and a
solitonic BS with $\sigma_0=0.2$ and central scalar amplitude
$A=0.157$, using resolutions $\Delta x_l = M/32$,
$\Delta x_m=M/48$ and $\Delta x_h = M/64$. In
Fig.~\ref{fig:E_convergence}, we show the resulting convergence
properties for the cumulative GW energy $E$ computed from
the Newman-Penrose scalar multipoles $\Psi^4_{\ell m}$ by
\begin{align}
    \dot{E}_{\ell m}(t) &= \frac{1}{16 \pi} \left|\int_{-\infty}^t \Psi^4_{\ell m} \du \tilde{t}   \right|^2, \\
    \dot{E} &= \sum_{\ell, m} \dot{E}_{\ell m}.
\end{align}

\begin{figure*}[hbt]
    \includegraphics[width=\linewidth]{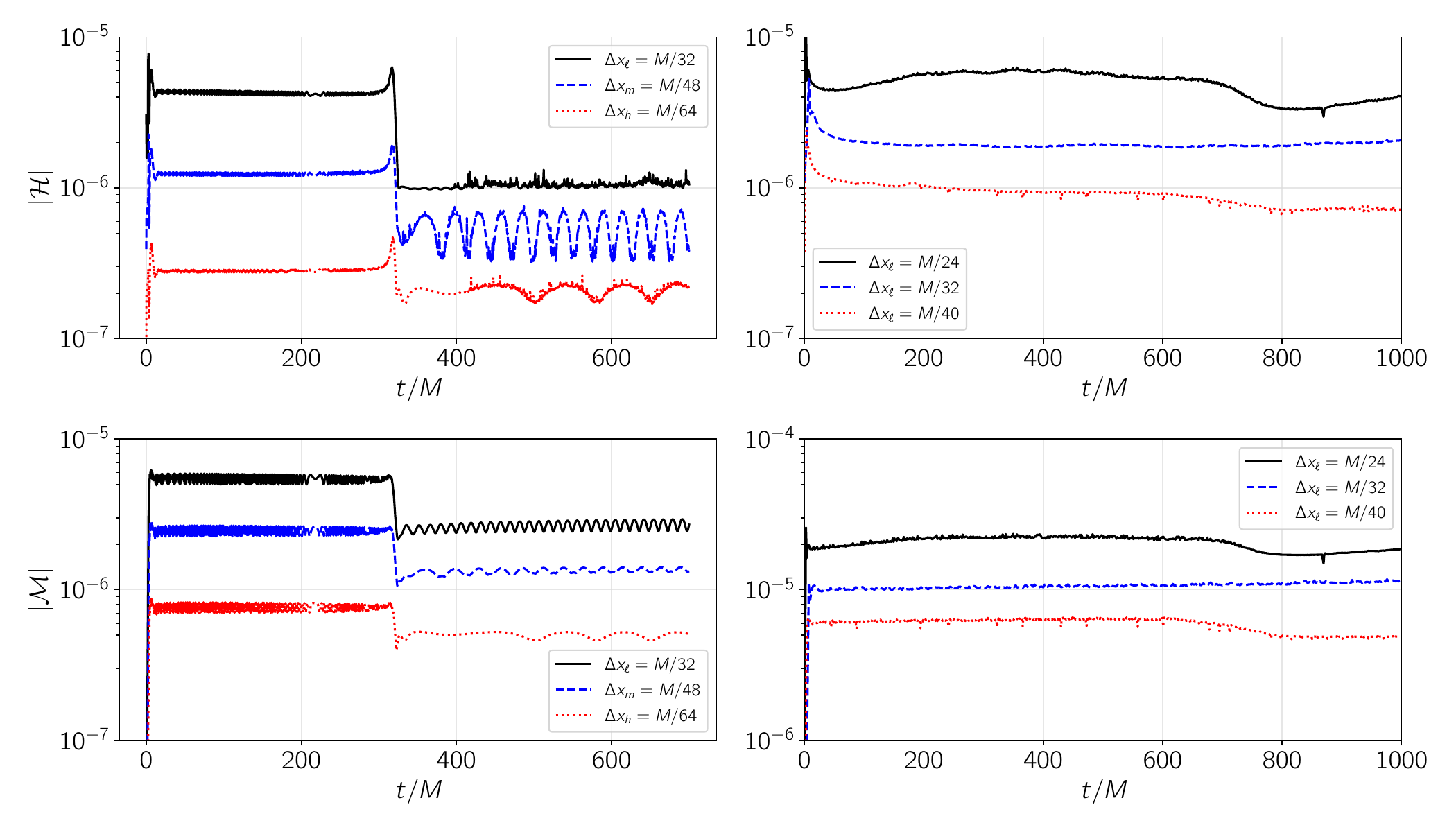}
    \caption{ $L^2$ norms of the Hamiltonian (top) and magnitude of the momentum (bottom) constraints at different resolutions $\Delta x$ over time, for two configurations: (left) an equal-mass, head-on BS-BH collision using a solitonic BS with $\sigma_0 = 0.2$ and central amplitude $A_0 = 0.157,$ and (right) our mini BS inspiral configuration (see Table~\ref{tab:inspiral_info}).
    }
    \label{fig:constraints}
\end{figure*}

The figure
demonstrates second-order convergence during the infall
stage which, starting at merger, gradually rises to between third and fourth order at later times. This observation is compatible with
the discretization schemes employed in \textsc{grchombo},
fourth-order finite differencing in space and time
combined with second-order accurate interpolation at the refinement boundaries. The latter are more prominent during
the dynamical infall and plunge stage which involve frequent
regridding, whereas the mesh refinement remains static
post-merger. The total GW energies obtained at different
resolutions are $E_l/M= 6.11 \times 10^{-4} $, $E_m/M= 6.27 \times 10^{-4}$ and $E_h/M = 6.31 \times 10^{-4}$
corresponding to convergence between third and fourth order  as well as a discretization error
of $0.4\%$ for our production resolution $\Delta x_h=M/64$.

A corresponding analysis of the dominant $\ell=m=2$
multipole of 
the Newman-Penrose scalar yields the same convergence and a time-dependent error of up to $1.5\%$. 
We have repeated this convergence analysis for other binaries
with larger boost velocities and unequal mass ratios. These 
yield larger discretization errors up to $1.5\%$ for our
most demanding configurations. 

To estimate the uncertainties arising from GW extraction
at finite radius, our simulations employ six equally-spaced extraction radii between $R_{\rm ex} = 80M$ and $R_{\rm ex} = 140M$. We extrapolate those values to infinity by using a first-order fit in $1 / R$ as described in Ref.~\cite{Radia_2021}.
From this analysis we obtain a numerical uncertainty
$\sim 1\%$ for $R_{\rm ex}=140 M$ 
independent of the binary parameters $v$
and $q$. Thus, the total error budget for our head-on GW energies is $1.5\%$ to $3\%$ for the configurations considered in this work, the larger (smaller) estimate aplying to
high (low) boost velocities and equal (unequal) mass ratios.

Finally, Fig.~\ref{fig:constraints} shows the convergence of the Hamiltonian and momentum constraints for one representative head-on merger and one inspiral configuration.
Overall, we see that our results are consistent with convergence between second and fourth order in both cases.

\end{document}